\documentclass[iicol,sn-aps]{sn-jnl}% American Physical Society (APS) Reference Style
\usepackage{natbib}
\usepackage{physics}
\usepackage[utf8]{inputenc}
\usepackage[english]{babel}
\usepackage{amsmath}
\usepackage{amsfonts}
\usepackage{amssymb}
\usepackage{graphicx}
\usepackage{xcolor}
\usepackage[T1]{fontenc}
\usepackage{mathrsfs}
\usepackage{braket}
\usepackage{geometry}
\usepackage{here}
\usepackage{lmodern}
\usepackage{array}
\usepackage{url}
\usepackage{textcase}
\usepackage{bm}
\usepackage{fancyvrb}
\jyear{2021}%

\theoremstyle{thmstyleone}%
\theoremstyle{thmstyletwo}%

\theoremstyle{thmstylethree}%

\usepackage{xcolor}
\usepackage{ifthen}
\def\Version{R}% Choix de la version a compiler (R (révision) ou O (originale))
\newcommand{\revision}[2]{%
	\ifthenelse{
		\equal{\Version}{R}
	}{{\color{black} #1}}{#2}
}

\begin{document}

\title[Article Title]{Survey of the Hierarchical Equations of Motion in Tensor-Train format for non-Markovian quantum dynamics}


\author[1]{\fnm{Etienne} \sur{Mangaud}}\email{etienne.mangaud@univ-eiffel.fr}

\author[2]{\fnm{Amine} \sur{Jaouadi}}\email{ajaouadi@ece.fr}

\author[3]{\fnm{Alex} \sur{Chin}}\email{alex.chin@insp.upmc.fr}

\author*[4]{\fnm{Mich\`ele} \sur{Desouter-Lecomte}}\email{michele.desouter-lecomte@universite-paris-saclay.fr}

\affil[1]{\orgdiv{MSME}, \orgname{Universit\'e Gustave Eiffel, UPEC, CNRS}, \orgaddress{\city{Marne-La-Vall\'ee}, \postcode{F-77454}, \country{France}}}

\affil[2]{\orgdiv{ECE - Paris}, \orgname{Graduate School of Engineering}, \orgaddress{\city{Paris}, \postcode{75015}, \country{France}}}

\affil[3]{\orgdiv{Institut des Nanosciences de Paris}, \orgname{Sorbonne Université, CNRS}, \orgaddress{\city{Paris}, \postcode{75015}, \country{France}}}

\affil*[4]{\orgdiv{Institut de Chimie Physique UMR8000}, \orgname{Universit\'e Paris-Saclay, CNRS}, \orgaddress{ \city{Orsay}, \postcode{91405}, \country{France}}}


\abstract{This work is a pedagogical survey about the hierarchical equations of motion and their implementation with the tensor-train format. These equations are a great standard in non-perturbative non-Markovian open quantum systems. They are exact for harmonic baths in the limit of relevant truncation of the hierarchy. We recall the link with the perturbative second order time convolution equations also known as the Bloch-Redfield equations.  \textcolor{black}{ Some theoretical tools characterizing non-Markovian dynamics such as the non-Markovianity measures or the dynamical map are also briefly discussed in the context of HEOM simulations}. The main points of the tensor-train expansion are illustrated in an example with a qubit interacting with a bath described by a Lorentzian spectral density. Finally, we give \textcolor{black}{three} illustrative applications in which the system-bath coupling operator is similar to that of the analytical treatment. The first example revisits a model in which population-to-coherence transfer via the bath creates a long-lasting coherence between two states. The second one is devoted to the computation of stationary absorption and emission spectra. We illustrate the link between the spectral density and the Stokes shift in situations with and without nonadiabatic interaction. \textcolor{black}{Finally, we simulate an excitation transfer when the spectral density is discretized by undamped modes to illustrate a situation in which the TT formulation is more efficient than the standard one.} }




\maketitle

\section{Introduction}
\label{secion:Intro}
Simulating quantum dynamics of complex systems with a large number of degrees of freedom (DoF) remains a computational challenge. However, measured observables often depend on a limited number of DoFs. Thus, the full system can be described as an active subsystem embedded in an environment, which makes fluctuate the energy levels of the subsystem. The latter is often described quantum mechanically while the extended environment is treated by a wide range of possibilities based on semi-classical or quantum or statistical approaches adequately chosen with respect to the choice of system-environment partitioning. When the number of DoFs increases, standard methods become computationnally untractable,  which is known as the "curse of dimensionality". In this context, the low-rank tensor decomposition has aroused a constantly growing interest. When the surrounding is modelled by an ensemble of discrete modes, Multi Configuration Time-Dependent Hartree (MCTDH) and the multi-layer version (ML-MCTDH) \cite{Manthe2008, Thoss2009, Meyer2011,Reichman2021} are based on tensor network algorithms, mainly the Tucker and hierarchical Tucker tensors \cite{Bader2009,Tobler2013}. Similarly, the expression of the time-dependent multi-mode wave functions in Matrix Product State (MPS) also called Tensor Train  (TT) \cite{Oseledets2011,Oseledets2015, Oseledets2016,Orus2014} expansion has revealed its efficiency in many applications \cite{Chin2016,SchroderChin2019,Chin2019,Dunnett2021,Dunnett2021bis, Baiardi2019,Ma2019,Plenio2019,Batista2022,Schmidt2022}.

In the usual approach of open quantum systems based on statistical mechanics with a surrounding at thermal equilibrium, the environment is described less explicitly and the active system is treated by a reduced density matrix by tracing over the bath degrees of freedom. Path integrals derived from the Feynman–Vernon influence functional \cite{Feynman1963} and the hierarchical equations of motion (HEOM) \cite{Kubo1989,Tanimura2006,Tanimura_rev_20} are a priori exact methods for harmonic baths and are closely related \cite{Yan2007,Yan2009}. HEOM may also be derived from the Nakajima–Zwanzig \cite{Nakajima1958,Zwanzig1960} partition of the Liouville equation by using the cumulant expansion of the reduced propagator and properties of the Gaussian distribution of the bath linked to the harmonic approximation \cite{ Ishizaki_Fleming2009}.  Both Path Integral and HEOM formalisms have also been recently treated by the tensor formalism in the Tensor Network Path Integral \cite{Walters2022} and in the MPS \cite{Shi2018,Shi2020,Borrelli2021,Borrelli_Gelin2021} or Tucker and hierarchical Tucker tensor format for HEOM \cite{Shi2021}. HEOM has been applied to describe many \textcolor{black}{physico-chemical} processes (see the recent review by Y. Tanimura \cite{Tanimura_rev_20}), for instance, excitation transfer in photosynthetical complexes \cite{Ishizaki_Fleming2009,Fleming2010,Kreisbeck2012,Fleming2015,Zhao2015,Limmer2022} or in other devices \cite{Tanimura2021}, non-adiabatic interactions and electron transfer \cite{Tanaka_Tanimura2010,MangaudDMP2015,Thiago2016}, dynamics via conical intersections \cite{Gelin2016,Duan2017,Dijkstra2017,MDL_19,Breuil2021,Jaouadi2022}, proton tranfer \cite{BorelliTanimura2020}, laser optimal control \cite{MDL2018,Jaouadi2022}, non-equilibrium fluxes \cite{Kato2016,Shi2017,Thoss2021}, \textcolor{black}{and} non-linear spectroscopies \cite{Tanimura_rev_20,Fleming2019,Zhu2011,Tanimura2012}.

It is worth noting that \textcolor{black}{accounting for temperature is handled in a similar way in the HEOM system of equations with discrete undamped modes \cite{Shi2014} and in a recent MPS implementation with wave functions \cite{ fchem2021} in the context of T-TEDOPA formalism \cite{Tamascelli2019} }. \textcolor{black}{In the latter,} the wave function approach \textcolor{black}{uses} discrete vibrational bath transformed into a chain and introduces finite temperature by sampling two baths representing absorption and emission respectively, with the latter being described by a bath of oscillators with negative frequencies. Emission into the vacuum of these negative modes mimics the absorption of environmental quanta that would be present in a physical (mixed-state) thermal bath, allowing a pure wavefunction description to capture the physics of a mixed-state initial condition without the need for thermal sampling \cite{Tamascelli2019, fchem2021}. On the other hand, the particular implementation of HEOM with undamped discrete modes also samples baths with positive and negative frequencies \cite{Shi2014}. By comparing \textcolor{black}{these} methods, we also emphasize that even if the reduced density matrix in open quantum systems is obtained by tracing over the bath modes, this does not mean that the information about the environment disappears. Relevant information about the baths may be extracted, for instance, the time-dependent distribution of the collective bath modes in each electronic state, which may be seen as the square modulus of a dissipative wave packet \cite{Shi2012,Shi2014,ChinChevet2019,MDL_19} or projection of the coherence among electronic states along the collective modes \cite{MDL_19} and fluxes \cite{Shi2017}. 

In this work, we present a pedagogical survey about HEOM and their implementation with the TT format. We recall the main lines and in particular the link with the perturbative second order time convolution equations also known as the Bloch-Redfield equations, which are an illuminating step to understand the complicated structure of HEOM. \textcolor{black}{We briefly discuss how HEOM may be used to compute some theoretical tools (non-Markovianity measure or dynamical map) related to non-Markovian dynamics. } We then present the principal points of the TT expansion by giving the expressions related to a simple example where a qubit interacts with a bath. Finally, we give \textcolor{black}{three} applications based on models in which the system-bath coupling operator is similar to the one used in the survey. \textcolor{black}{Finally, the appendix explains how to encode the main expressions with a Python package.}

\section{Open quantum systems and HEOM}
\label{section:OQS_HEOM}

The standard starting point in open quantum system \cite{Breuer2002,Weiss2012,Kuhn2011} is the partition of the DoFs of the full complex system. The active subsystem may be only electronic DoFs like in the usual spin-boson model or include some Brownian coordinates coupled to residual baths \cite{Chenel2014,MangaudDMP2015,Nazir2016,Gelin2016}. The generic partitioning of the full Hamiltonian in three parts is written \textcolor{black}{as}:
\begin{equation}
H={{H}_{S}}+{{H}_{SB}}+{{H}_{B}}
\label{eq:partH}
\end{equation}
where ${H}_{S}$ and ${H}_{B}$ are the Hamiltonians of the active system and of the vibrational or phonon bath(s) respectively. The system Hamiltonian may be time dependent if it contains interaction with \textcolor{black}{external} fields. When the system interacts with $N_{bath}$, the system-bath coupling is ${{H}_{SB}}=\sum\nolimits_{\alpha =1}^{{{N}_{bath}}}{{{S}_{\alpha }}}{{B}_{\alpha }}$. \textcolor{black}{${S}_{\alpha }$} and ${{B}_{\alpha }}$ are operators in the space of the system and in the complementary space respectively.

In the electronic-nuclear partition, the system operators $S_{\alpha}$ are $n\times n$ matrices with  $n$ the number of electronic states. \textcolor{black}{They act as projectors on some states when the baths tune the electronic energies (i.e. are diagonally coupled) or transition matrices between some of them when the baths make fluctuate the off-diagonal electronic coupling such as in the case of conical intersections} \cite{Duan2017,Dijkstra2017,MDL_19,Breuil2021,Jaouadi2022}. \textcolor{black}{In other system-bath partition cases}, the system operators are the Brownian coordinates included in the active subspace \cite{Chenel2014,MangaudDMP2015,Nazir2016,Gelin2016}. \textcolor{black}{Furthermore,}  each bath operator $B_{\alpha}$ is a linear combination of the position operators $q_j$ of the oscillators ${{B}_{\alpha }}=\sum\nolimits_{j}^{{{N}_{\alpha }}}{c_{j}^{(\alpha )}}{{q}_{j}}$ in the discrete bath representation with $N$ oscillators. The expression of the coupling coefficients  $c_{j}^{(\alpha )}$ depends on the partition and on the choice of the coordinates. \textcolor{black}{I}n the following, we adopt mass-weighted coordinates and the electronic-nuclear partition. 

The initial total density operator $\rho_{tot}(0) = \rho_S(0) \otimes \rho_{B,eq}$ is assumed to be factorized and is the product of the system density operator $\rho_S(0) $ and a thermally equilibrated bath density operator $\rho_{B,eq}=e^{-\beta H_B}/Tr_B\left[e^{-\beta H_B}\right]$. Extension to correlated initial conditions have been proposed \textcolor{black}{in Refs.}  \cite{Pomyalov2010,Dijkstra2010,Tanimura2014,Shi2015}. 

\subsection{Second order auxiliary operators }
It is very instructive to first examine the non-Markovian perturbative equations, which contain all the crucial tools occurring in HEOM. To do so, we consider the simplest case with only one bath, i.e. ${H}_{SB}=SB$. The exact formal Nakajima-Zwanzig equation given the evolution of the reduced density matrix $\dot{\rho_S }(t)=-\frac{i}{\hbar }T{{r}_{B}}\left[ H,{{\rho }_{tot}}(t) \right]$ may be written as:
\begin{equation}
\dot{\rho_S }(t)={{L}_{S}}\rho_S (t)+\int_{0}^{t}{K(t,\tau )\rho_S (\tau )d\tau } + I(t)
\label{eq:NZeq}
\end{equation}
with ${{L}_{S}}\centerdot =-i\left[ {{H}_{S}},\centerdot  \right]$ the system Liouvillian and $\hbar =1$. $K(t,\tau)$ is the memory kernel which embeds the bath influence on the system and $I(t)$ is an initial correlation term which cancels when the system and bath can be initially factorized (i.e. $I(t) = 0$ when $\rho_{tot}(0) = \rho_S \otimes \rho_{B,eq}$).

At the second order in the ${{H}_{SB}}$ coupling, the memory kernel becomes \citep{Meier1999}\textcolor{black}{:}
\begin{align}
&\int_{0}^{t}{d\tau }K(t,\tau ){{\rho }_{S}}(\tau )= \int_{0}^{t}{d\tau }K^{(2)}(t, \tau ){{\rho }_{S}}(\tau )
 \nonumber \\
&= i\int_{0}^{t}{d\tau }\left[ S,\left\{ iC(t-\tau ){{U}_{S}}(t-\tau )S{{\rho }_{S}}(\tau ) \right. \right.
\nonumber \\
& \left. \left. \times U_{S}^{\dagger }(t-\tau ) \right\}+\left\{ hc \right\} \right]
\label{align:secondorder}
\end{align}
where $C(t)$ is the correlation function of the collective bath mode $C(t)={{\left\langle B(t)B(0) \right\rangle }_{eq}}$, $B(t)$ is the Heisenberg representation of the operator with Hamiltonian $H_B$ and ${{\left\langle \centerdot  \right\rangle }_{eq}}$ denotes the average over a Boltzmann distribution at \textcolor{black}{a given} temperature $T$. $U_S(t) = e^{-iH_S t}$ is the propagator of the system. The treatment up to the fourth order may be found in Ref. \cite{Cao2002} and \textcolor{black}{the} extension by the generalized master equation method in Ref. \cite{Geva2019}.

The quantum bath correlation function is the main descriptor of the bath \cite{Mukamel1995}. The usual step to go towards the second order auxiliary density operator (ADO) or HEOM is the representation of the correlation function as a sum of damped decaying functions:
\begin{equation}
C\left( t-\tau  \right)={{\left\langle B(t)B(\tau ) \right\rangle }_{eq}}=\underset{k=1}{\overset{K}{\mathop \sum }}\,{{\alpha }_{k}}{{e}^{i{{\gamma }_{k}}\left( t-\tau  \right)}}
\label{eq:C(t)}
\end{equation}
where ${{\alpha }_{k}}$ and ${{\gamma }_{k}}$ are complex \textcolor{black}{parameters}. The sum is a priori infinite but truncated to $K$ modes, which have been called the "artificial bath modes" in Ref. \cite{Meier1999}.  Some extensions to different analytical forms or arbitrary correlation functions have been proposed recently \textcolor{black}{in Refs.} \cite{Kleinekat2019,Yan2019,Ikeda2020,Yan2022,Lambert2020}. 

By inserting Eq.(\ref{eq:C(t)}) in Eqs.(\ref{eq:NZeq}) and (\ref{align:secondorder}), each artificial mode corresponds to a particular memory integral of the time dependent integro-differential equation. Each integral is set equal to an ADO :
\begin{align}
&{\rho }_{k}(t)/(i {\alpha }_{k})  \nonumber \\
&= \int_{0}^{t}{d\tau }\left[ S,\left\{ i{{e}^{-{{\gamma }_{k}}(t-\tau )}}{{U}_{S}}(t-\tau )S{{\rho }_{S}}(\tau )
\nonumber  \right. \right.\\
&\left. \left.\times U_{S}^{\dagger }(t-\tau ) \right\}+\left\{ hc \right\} \right].
\label{align:ADO}
\end{align}

The dimension of ${{\rho }_{k}}(t)$ is that of the $S$ operator and thus of ${{H}_{S}}$. A time local system of coupled equations may be obtained by taking the first derivative \cite{Meier1999,Kleinekat2004,Pomyalov2010}. Different choices are possible to define the ADOs. From Eq.(\ref{align:ADO}), we obtain the operational equations \cite{ Pomyalov2010}:
\begin{align}
&{{\dot{\rho }}_{S}}(t)={{L}_{S}}{{\rho }_{S}}(t)+i\underset{k=1}{\overset{K}{\mathop \sum }}\,{\left[ S,{{\rho }_{k}}(t) \right]}
 \nonumber \\
&{{\dot{\rho }}_{k}}(t)=\left( i{{\gamma }_{k}}+{{L}_{S}} \right){{\rho }_{k}}(t)+i\left[ {{\alpha }_{k}}S{{\rho }_{S}}(t) \right.
\nonumber \\
&\left. -{{{\tilde{\alpha }}}_{k}}{{\rho }_{S}}(t)S \right]
\label{align:eqcousecorder}
\end{align}
where ${{L}_{S}}$ is the system Liouvillian. The ${{\tilde{\alpha }}_{k}}$ parameters will be discussed below. 

\begin{figure}
 \centering
\includegraphics[width =1.\columnwidth,height=4cm]{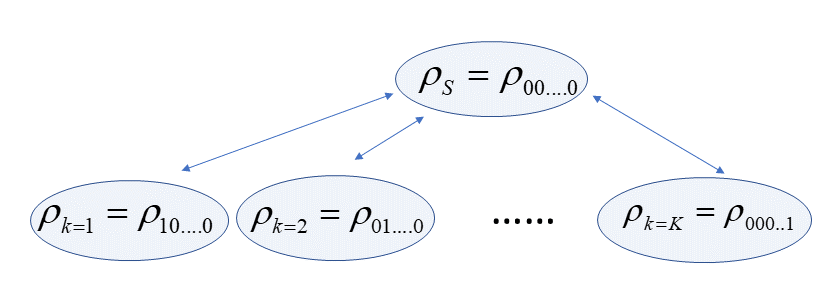}
\caption{Schematic representation of the auxiliary operators associated to each artificial bath mode $k = 1,K$ and involved in the second order master equation (Eq.( \ref{align:eqcousecorder}). Each ADO corresponds to a single excitation in the mode.  }
\label{fig:schemasecorder}
\end{figure}

Each ADO being associated \textcolor{black}{only} to one decay mode, it may be considered as resulting from a single excitation in this mode and denoted by an array of $K$ indexes with one in the $k^{th}$ position and zero everywhere else as shown in figure \ref{fig:schemasecorder}.  These ADOs will constitute the first level of the HEOM hierarchy. 

A real classical correlation function may be obtained by molecular dynamics \cite{Valleau2012,Fleming2015,Thiago2016,Dunnett2021bis} or directly from experimental results \citep{Marcus2002} and corrected to get the complex quantum correlation function satisfying the fluctuation-dissipation theorem \citep{Mukamel1995}. Direct parametrization of $C(t)$ \textcolor{black}{ fitted by different approaches} \cite{Lambert2020}, Prony method \cite{Yan2022} or expansion on Chebyshev or Bessel functions \cite{Kleinekat2019} have also been proposed. However, the relation with the bath spectral density, \textcolor{black}{defined as follows:} 
\begin{equation} 
C(t)=\int_{-\infty }^{\infty }{d\omega {{J}_{{}}}\left( \omega  \right){{\left( {{e}^{\beta \omega }}-1 \right)}^{-1}}}{{e}^{-i\omega t}}
\label{eq:CtoJ}
\end{equation}
where $\beta =1/{{k}_{B}}T$ is the Boltzmann factor, is currently used and the $\left\{ {{\alpha }_{k}},{{\gamma }_{k}} \right\}$ parameters are obtained from the parametrization of $C(\omega) = {{J}_{{}}}\left( \omega  \right){{\left( {{e}^{\beta \omega }}-1 \right)}^{-1}} $ where 
${{J}_{{}}}\left( \omega  \right)$ is independent of the temperature and ${{\left( {{e}^{\beta \omega }}-1 \right)}^{-1}}$ is the Bose function related to the quantum fluctuation-dissipation theorem. In the discrete case, the spectral density is defined from the system-bath coupling coefficients ${{c}_{j}}$ \textcolor{black}{related to $N_{disc}$ discrete modes}
\begin{equation}
J(\omega )=\frac{\pi }{2}\underset{j=1}{\overset{N_{disc}}{\mathop \sum }}\,{\frac{c_{j}^{2}}{{{\omega }_{j}}}}\delta (\omega -{{\omega }_{j}}).
\label{eq:J(w)}
\end{equation}

In the continuous representation, the spectral density is often approximated by an Ohmic function ${{J}_{Ohm}}\approx c\omega {{f}_{cutoff}}(\omega )$ or a super Ohmic function ${{J}_{SuperOhm}}\approx c{{\omega }^{3}}{{f}_{cutoff}}(\omega )$, with $c$ a constant and $f_{cutoff}$ an exponential or Lorentzian cutoff. The Ohmic function is used for solvents while the super Ohmic one is relevant for the solid phase and phonon baths. In order to get a parametrization of the bath correlation function (Eq.(\ref{eq:C(t)})) through Eq.(\ref{eq:CtoJ}), one has to perform a fitting procedure of the spectral functions with relevant analytical functions. A special care in fitting the spectral density low frequency behavior is often necessary to get accurate results \citep{Eisfeld2014}. Here, we will discuss only the cases of Ohmic and super-Ohmic Lorentzian functions.  We adopt the Tannor-Meier parametrization \cite{Meier1999} in which the  spectral density is fitted by \textcolor{black}{ ${n}_{lor}$} two-pole Lorentzian functions depending on three parameters $\left( {{p}_{l}},{{\Omega }_{l}},{{\Gamma }_{l}} \right)$ \textcolor{black}{where ${{p}_{l}}$ is the associated  weight, $ {\Omega}_{l}$ the central frequency, and ${\Gamma }_{l}$ the bandwidth.} 
\begin{equation}
{{J}_{Ohm}}\left( \omega  \right)=\sum\limits_{l=1}^{{{n}_{lor}}}{\frac{{{p}_{l}}\omega }{\Upsilon \left( {{\Omega }_{l}},{{\Gamma }_{l}} \right)}}
\label{eq:Johmic}
\end{equation}
with 
\begin{equation}
{\Upsilon \left( {{\Omega }_{l}},{{\Gamma }_{l}} \right)=\left[ {{\left( \omega +{{\Omega }_{l}} \right)}^{2}}+\Gamma _{l}^{2} \right]\left[ {{\left( \omega -{{\Omega }_{l}} \right)}^{2}}+\Gamma _{l}^{2} \right] \nonumber}
\end{equation}
All the $\left\{ {{\alpha }_{k}},{{\gamma }_{k}} \right\}$ parameters may be obtained from the integral (\ref{eq:CtoJ}) after substitution of Eq.(\ref{eq:Johmic}). Each Lorentzian corresponds to two artificial decay modes. The Bose function generates an infinite series of terms coming from its poles. They are known as the Matsubara terms. In practice, the number of Matsubara terms is very small at high temperature but may become numerous at low temperature rendering the ADO method computationally more demanding. In the low temperature regime, Padé approximants of the Bose function \citep{Yan2011}, \textcolor{black}{ fitting procedure of $C(t)$ used to capture the Matsubara terms} \cite{Lambert2019}  or a recent correction scheme \cite{Fay2022} can be used. \textcolor{black}{The analytical expressions of $\left\{ {{\alpha }_{k}},{{\gamma }_{k}} \right\}$ (Eq.(\ref{eq:C(t)})) as functions of the $\left( {{p}_{l}},{{\Omega }_{l}},{{\Gamma }_{l}} \right)$ parameters of $J_{Ohm}(\omega)$ and those related to the Matsubara terms} are  given in Refs. \citep{Pomyalov2010} or \citep{Chenel2014}. One also \textcolor{black}{could} find in these references the expression of the ${{\tilde{\alpha }}_{k}}$ \textcolor{black}{parameters} (see Eq.(\ref{align:eqcousecorder})). They come from a particular expression of the complex conjugate of $C(t)$ that may be written \textcolor{black}{as:} ${{C}^{*}}\left( t \right)=\underset{k=1}{\overset{K}{\mathop \sum }}\,{{\tilde{\alpha }}_{k}}{{e}^{i{{\gamma }_{k}}t}}$ with the same ${{\gamma }_{k}}$ as in Eq.(\ref{eq:C(t)}). 
In the super Ohmic case, the fitting functions have four poles leading to four artificial decay channels
\begin{equation}
{{J}_{SuperOhm}}\left( \omega  \right)=\sum\limits_{l=1}^{{{n}_{lor}}}{\frac{{{p}_{l}}{{\omega }^{3}}}{\Upsilon \left( {{\Omega }_{l,1}},{{\Gamma }_{l,1}} \right)\Upsilon \left( {{\Omega }_{l,2}},{{\Gamma }_{l,2}} \right)}}.
\label{eq:Jsuperohmic}
\end{equation}
The analytical expressions to get the $\left\{ {{\alpha }_{k}},{{\gamma }_{k}} \right\}$ parameters of $C(t)$ (Eq.(\ref{eq:C(t)})) from $\left( {{p}_{l}},{{\Omega }_{l,1}},{{\Gamma }_{l,1}},{{\Omega }_{l,2}},{{\Gamma }_{l,2}}\right)$ parameters of $J_{SuperOhm}(\omega)$ are gathered in Ref. \cite{MangaudMeier2017}.

\subsection{HEOM}
First, we discuss the case with a single bath, i.e. \textcolor{black}{we consider} only one $S$  operator. The generalization \textcolor{black}{will be} discussed at the end of the section.  When the bath correlation time becomes long with respect to the system characteristic timescale due to a strongly peaked spectral density, the high non-Markovianity generally involves a non-perturbative regime. HEOM are one of the reference dynamical methods for open quantum systems modelled with a harmonic bath. The equations giving the evolution of the reduced density matrix were originally derived for a Drude-Lorentz spectral density in the high-temperature limit from the Kubo stochastic Liouville equation \citep{kubo_stochastic_1963,Kubo1989,Tanimura2006, Tanimura_rev_20} and the Feynman-Vernon influence functional formalism \cite{Yan2007,Yan2009}. Another approach to derive these equations is by referring to the remarkable property that the cumulant expansion limited to second order is exact when the bath statistics are Gaussian \citep{Ishizaki_2005,Ishizaki_Fleming2009}. This is based on the Wick theorem \cite{Wick1950}. The reduced density matrix ${{\rho }_{S,I}}(t)$  for the system in interaction representation $\rho_{S,I}(t) = e^{i H_0 t} \rho_S(t) e^{-i H_0 t}$ with $H_0 = H_S + H_B$, is given by the partial trace over the bath of the time evolution of the total density matrix \textcolor{black}{:}
\begin{equation}
{{\rho }_{S,I}}(t)=T{{r}_{B}}\left[ {\mathcal{T}^{(+)}}{{e}^{\int\limits_{0}^{t}{d\tau {{L}_{SB,I}}(\tau )}}}\rho _{B}^{eq} \right]{{\rho }_{S,I}}(0)  
\label{eq:rho_sI}
\end{equation}
where $\mathcal{T}^{(+)}$ is a time-ordering operator and a factorization is assumed at the initial time $t = 0$.  ${{L}_{SB,I}}(t)$ is the Liouvillian in interaction representation, ${{L}_{SB,I}}(t)\centerdot =-i[{{S}}(t){{B}}(t),\centerdot ]$, where $S(t)={{e}^{i{{H}_{S}}t}}S{{e}^{-i{{H}_{S}}t}}$ and $B(t)={{e}^{i{{H}_{B}}t}}S{{e}^{-i{{H}_{B}}t}}$.

At the second order in the cumulant expansion, Eq.(\ref{eq:rho_sI}) becomes 
\begin{equation}
{{\rho }_{S,I}}(t)={\mathcal{T}^{(+)}}{{e}^{\int\limits_{0}^{t}{d\tau \mathcal{K}_I(\tau )}}}{{\rho }_{S,I}}(0).
\label{eq:rho_sIsecorder}
\end{equation}
$\mathcal{K}_I(\tau) = \int\limits_{0}^{\tau} dt'K_I^{(2)} (\tau, t')$ corresponds to the second-order memory term occurring in the second-order perturbation theory (Eq.(\ref{align:secondorder})) but here, the second order is an exact expression for the cumulant expansion : 
\begin{align}
& \int_0^\tau dt' K_I^{(2)} (\tau,t') \centerdot
  \nonumber \\
& = -  [S (\tau), \int\limits_{0}^{\tau} dt'   C(\tau-t') S(t')   \centerdot   - \lbrace h.c. \rbrace ].
 \label{eq:K_ordre2}
\end{align}
By inserting the correlation function parametrization (Eq.(\ref{eq:C(t)})), equation (\ref{eq:K_ordre2}) is separable into a sum of $K$ operators $\mathcal{K}_{Ik}$ where $K$ is the number of artificial decay modes. In addition, each $\mathcal{K}_{Ik}$ is decomposed as :
\begin{equation}
  \mathcal{K}_{Ik} (\tau) = \Phi_I(\tau) W_{Ik}(\tau)
\label{eq:K_k}
\end{equation}
where 
\begin{equation}
\Phi_I(\tau) \centerdot = -[S(\tau), \centerdot]
\label{eq:Phi}
\end{equation}
and 
\begin{equation}
W_{Ik} (\tau) = \int\limits_{0}^{\tau} dt' e^{i{{\gamma }_{k}}\left( \tau - t' \right)}{{\Theta }_{Ik}}\left( t' \right)
\label{eq:W_k}
\end{equation}
with
\begin{equation}
\Theta_{Ik}(t') \centerdot = {{\alpha }_{k}}S\left( t' \right)\centerdot - {{{\tilde \alpha }_{k}}}\centerdot S\left( t'  \right). 
\label{eq:Theta_k}
\end{equation}

\begin{figure}
 \centering
\includegraphics[width =1.\columnwidth,height=5cm]{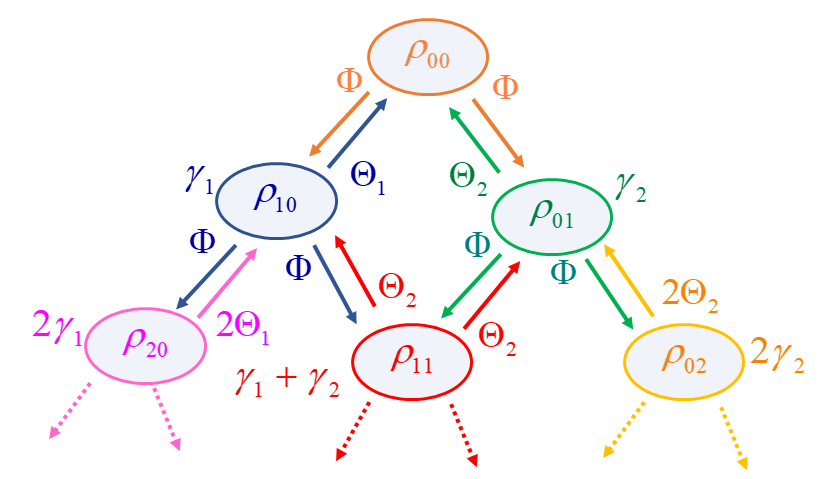}
\caption{Schematic representation of the auxiliary operators of the HEOM hierarchy up to level $L=2$ for the case with two artificial bath modes $K=2$. $\Phi$ acts on matrices for which one occupation number increases by one unit. $\Theta_k$ is applied on matrices for which one occupation number decreases by one unit. The sum of the decay rates $n_k\gamma_k$ multiplying each matrix is indicated. The schematic representation of the terms involved in the derivative of one matrix is complete for $\rho_{00}$ (orange); $\rho_{10}$ (blue) and $\rho_{01}$ (green).   }
\label{fig:schemaHEOM}
\end{figure}

The exact solution takes the \textcolor{black}{following} form:
\begin{align}
 & {{\rho }_{S,I}}(t)={\mathcal{T}^{(+)}}\prod\nolimits_{k}^{K}{{{e}^{\int_{0}^{t}{d\tau }{\mathcal{K}_{Ik}}(\tau )}}}{{\rho }_{S,I}}(0) 
 \nonumber \\
 & ={\mathcal{T}^{(+)}}\prod\nolimits_{k}^{K}{{{e}^{\int_{0}^{t}{d\tau }\int_{0}^{\tau }{dt'\Phi_I (t'){{e}^{i{{\gamma }_{k}}(\tau -t')}}{{\Theta }_{Ik}}(t')}}}}
 \nonumber \\
 &\times {{\rho }_{S,I}}(0).
 \label{eq:rhoformel}  
\end{align}
The master equation can then be derived \textcolor{black}{as}
\begin{equation}
{{\dot{\rho }}_{S,I}}(t)={{\mathcal{T}}^{(+)}}\sum\nolimits_{k}^{K}{{{\mathcal{K}}_{Ik}}(t){{\rho }_{S,I}}(t)}.
\label{eq:drhoformel}
\end{equation}
This equation is time non-local since $\mathcal{K}_{Ik}(t)$ contains an integral. A time local system of coupled equations is obtained by defining the ADOs by the same strategy as in the second order approach :
\begin{align}
&{{\rho }_{\mathbf{n},I}}(t) \nonumber \\
&={\mathcal{T}^{(+)}}\prod\nolimits_{k}^{K}{{{W}_{Ik}}}{{(t)}^{{{n}_{k}}}}{{e}^{\int_{0}^{t}{d\tau {\mathcal{K}_{Ik}}(\tau )}}}{{\rho }_{S,I}}(0)
\label{eq:ADOHEOM}
\end{align}
where 
\begin{equation}
\mathbf{n}=\left\{ {{n}_{1}},...,{{n}_{k}},...{{n}_{K}} \right\} 
\label{eq:indicen}
\end{equation}
in a vector of nonnegative integers giving the occupation number in each artificial decay mode. The case $\mathbf{n}=\left\{ 0,...,0,...0 \right\}$ corresponds to $\rho_{S,I}$. Time local coupled equations among the ADOs may be derived by working with the Fourier-Laplace transforms of Eqs. (\ref{eq:rhoformel}) and (\ref{eq:ADOHEOM}) and using integration by parts \cite{Fleming2010,MDL_19}. One recovers the relations established in Ref.\cite{Kubo1989}. After the inverse Laplace-Fourier transform and the return in the Schr\"odinger representation, the HEOM read :
\begin{align}
{{{\dot{\rho }}}_{\mathbf{n}}}(t)=  & {{L}_{S}}{{\rho }_{\mathbf{n}}}(t)+i\sum\limits_{k=1}^{K}{{{n}_{k}}{{\gamma }_{k}}{{\rho }_{\mathbf{n}}}}(t)
 \nonumber \\
&-i\left[ S,\sum\limits_{k=1}^{K}{{{\rho }_{\mathbf{n}_{k}^{+}}}(t)} \right]
\nonumber \\
&-i\sum\limits_{k=1}^{K}{{{n}_{k}}\left( {{\alpha }_{k}}{{S}}{{\rho }_{\mathbf{n}_{k}^{-}}}(t)-{{{\tilde{\alpha }}}_{k}}{{\rho }_{\mathbf{n}_{k}^{-}}}(t)S \right)}
\label{align:HEOM} 
\end{align}
or more shortly by using definitions (\ref{eq:Phi}) and (\ref{eq:Theta_k}) in Schr\"odinger representation (i.e., $\Phi(\tau) \centerdot = -[S, \centerdot]$ and $\Theta_k \centerdot = {{\alpha }_{k}}S\centerdot - {{{\tilde \alpha }_{k}}}\centerdot S$):
\begin{align}
&{{\dot{\rho  }}_{\mathbf{n}}(t)}={{L}_{S}}{{\rho }_{\mathbf{n}}} \nonumber \\
&+i\sum\limits_{k=1}^{K}{\left( {{n}_{k}}{{\gamma }_{k}}{{\rho }_{\mathbf{n}}}-\Phi {{\rho }_{\mathbf{n}_{k}^{+}}}-{{n}_{k}}{{\Theta }_{k}}{{\rho }_{\mathbf{n}_{k}^{-}}} \right)}.
\label{eq:HEOMshort}
\end{align}
The subscripts
\begin{equation}
\mathbf{n}_{k}^{+}=\left\{ {{n}_{1}},\cdots ,{{n}_{k}}+1,\ldots ,{{n}_{{{n}_{K}}}} \right\} \nonumber 
\end{equation}
and 
\begin{equation}
\mathbf{n}_{k}^{-}=\left\{ {{n}_{1}},\cdots ,{{n}_{k}}-1,\ldots ,{{n}_{{{n}_{K}}}} \right\} \nonumber 
\end{equation}
denote the matrices for which one occupation number differs by one unit in the hierarchy ${{n}_{k}}\to {{n}_{k}}\pm 1$. The sum of the occupation numbers defines the level of the hierarchy $L=\sum n_k$. The total number of matrices when the hierarchy is limited at level $L$ is $(L+K)!/L!K!$. Each matrix is connected to \textcolor{black}{the} matrices of the lower or \textcolor{black}{the} upper level\textcolor{black}{s}. Figure \ref{fig:schemaHEOM} illustrates the case with two artificial decay channels (i.e. one Ohmic Lorentzian in the spectral density and no Matsubara term). Initially, $\rho_{00...0}=\rho_S$ and all the ADOs are zero. They represent the initial bath at equilibrium. When the final relaxed state is reached, the converged ADOs represent the new equilibrium and may be taken as the initial condition to describe another process affecting the equilibrated system\textcolor{black}{. F}or instance to describe the stationary fluorescence. All the terms occurring in the derivative of each ADO are schematized in figure \ref{fig:schemaHEOM} for a simple example with two bath modes. 

The generalization is straightforward when there are $N_{bath}$ uncorrelated baths, each associated with $K_b$ artificial bath modes. Then, $K=\sum\nolimits_{b=1}^{{{N}_{bath}}}{{{K}_{b}}}$. Each bath is linked to the system by an operator $S_b$ and the equations become :
\begin{align}
 &{{\dot{\rho }}_{\mathbf{n}}(t)}={{L}_{S}}{{\rho }_{\mathbf{n}}} +i\sum\limits_{b}^{{{N}_{bath}}}{\sum\limits_{k=1}^{{{K}_{b}}}{\left( {{n}_{b,k}}{{\gamma }_{b,k}}{{\rho }_{\mathbf{n}}}\right.}}
 \nonumber \\
 &{{\left. -{{\Phi }_{b}}{{\rho }_{\mathbf{n}_{k}^{+}}}-{{n}_{b,k}}{{\Theta }_{b,k}}{{\rho }_{\mathbf{n}_{k}^{-}}} \right)}}.
\label{eq:HEOMshortNb}
\end{align}

\textcolor{black}{An alternative to the continuous representation of the spectral density with $K$ artificial modes is the discretization with $N_{disc}$ undamped modes in the spirit of MCTDH or ML-MCTDH computations \cite{Burghardt2012,Burghardt2020,Burghardt2021,kuhn2016} or MPS simulations in the Hilbert space \cite{ SchroderChin2019,Chin2019,Dunnett2021,Dunnett2021bis}. Discretization has also been illustrated with HEOM \cite{Shi2014} and checked in TT format \cite{Shi2018,Shi2021}. The expression of the corresponding HEOM equations are given in Refs. \cite{Shi2014,Shi2018}.  The active system is then coupled to two identical baths with positive or negative frequencies describing emission and absorption of energy. Since there are two baths, the number of terms in the correlation function is large $K = 2 N_{disc}$ but there are no Matsubara terms associated with the poles of the Bose function in the continuous case.  The discrete couplings are smaller than those of the artificial modes. We will illustrate in one application that the TT implementation may be more efficient than the standard one in that case.}

The system of the coupled differential equations (\ref{eq:HEOMshortNb}) or those related to discrete modes given in refs. \cite{Shi2014,Shi2018} can then be solved using standard numerical methods such as Runge-Kutta 4 (RK4), Cash-Karp (RK4-5) adaptative step-size or Arnoldi algorithms. Although HEOM are an infinite system of differential equations, they are truncated at a maximum value of $n_k = n_{max}$. Thus, several simulations with increasing values of $n_{max}$ should be carried out until the results (density matrix or observables) are properly converged. \textcolor{black}{For comparison between the standard and TT formulations, we have used a home-made fortran code parallelized with OpenMP. Other implementations in python or on new architectures such as GPU \cite{ Kreisbeck2014} could be useful. See a review of different softwares in Ref. \cite{Lambert2020}.
}
\subsection{HEOM and non-Markovianity}
 \textcolor{black}{
  As mentioned above, HEOM are a standard method to tackle non-perturbative and non-Markovian regimes. Non-Markovianity obviously depends on the partition since it is roughly predicted by the characteristic timescales of the active system and the corresponding bath. These typical dynamical times may be very different with respect to the system definition. Inserting some effective bath coordinates into the system part (to reduce the coupling towards the residual bath and modify the corresponding timescales) may be an efficient strategy to change the Markovianity regime  \cite{Burghardt2010, Chenel2014, Nazir2014,MangaudDMP2015, Nazir2016, MDL_19}.  When the partition leads to a non-Markovian regime, an abundant literature has been devoted to the characterization of the non-Markovianity by different measures and from a fundamental perspective, to a mathematical definition of non-Markovian quantum dynamical maps, which is still an open problem. These fundamental questions are reviewed for instance in Refs. \cite{RevBreuer2016,deVega2017}. 
A detailed presentation of these items is beyond the scope of this paper and we summarize here only the main points.  A Markovian behavior is linked to a continuous loss of information from the open system to the surrounding while a flow from the environment back to the open system is the signature of a non-Markovian effect.  The return of information from the bath could modify the system dynamics. The measures aim at quantifying this backward flow. One may cite, among others, the trace distance measure \cite{Laine2009,RevBreuer2016}, the entanglement measure \cite{Rivas2010}, negativity of time-dependent canonical rates  \cite{Anderson2014}, and the Bloch volume measure \cite{Paternostro2013}. They are compared for instance in Refs. \cite{Anderson2014,Yi2015}. In the last case, the measure is based on an estimation of the volume of the accessible states in the generalized Bloch sphere for a $n-$state system. A non-monotonic decrease of this volume is a signature of non-Markovianity. Any density matrix may be expanded in the basis set of ${{n}^{2}}$ operators : the normalized identity $G_0=I/\sqrt{n}$ and $G_i (i = 1,...,n^2-1)$: the $n^2-1$ generators of $SU(n)$ \cite{Kimura2003,Aerts2014}. They are the Pauli matrices for $n = 2$ and the Gell-Mann matrices for $n = 3$. The volume of accessible states $V(t) =\det (\mathbf{F}(t))$ is obtained from the determinant of the matrix ${{F}_{m,n}}(t)=Tr\left( {{G}_{m}(0)} {{G}_{n}} (t)  \right)$. This requires $n^2$ propagations of the basis operators and this is easily obtained with HEOM \cite{ Atabek2017,Chevet2018,MDL2018,MangaudMeier2017, ChinChevet2019, MDL_19}. Similarly, the $n^2$ basis operators may be used to build the $n^2 \times n^2$ decoherence matrix \cite{Anderson2014,MangaudMeier2017}  ${{D}_{ij}}(t)=\sum\nolimits_{m=0}^{{{n}^{2}}-1}{Tr\left[ {{G}_{m}}{{G}_{i}}{{\Lambda }_{t}}\left[ {{G}_{m}(t)} \right]{{G}_{j}} \right]}$ where ${{\Lambda }_{t}}\left[ {{G}_{m}(t)} \right]$ is the right member of the master equation 
${{\dot{G}}_{m}}(t)={{\Lambda }_{t}}\left[ {{G}_{m}(t)} \right]$, in other words, the expression of ${{\dot{\rho }}_{00..0}}(t)$ when the initial state is $G_m$ (see Eq.\eqref{align:HEOM}). The eigenvalues of this decoherence matrix are called the canonical decay rates ${{\gamma }_{k}}(t)$  \cite{Anderson2014,MangaudMeier2017}. They witness non-Markovianity when some of them become negative signalizing an information return towards the system.}

 \textcolor{black}{The concept of dynamical map in the theory of open quantum is relevant when the initial state of the total system is a factorized product state but it remains a debated point for initial system-bath entangled state \cite{ RevBreuer2016, Aspuru2010, Rosario2013}. While the master equation is related to the time derivative ${{\dot{\rho}}}(t)={{\Lambda }_{t}}\left[ {{\rho}(t)} \right]$, the corresponding dynamical map $\phi_{t}$ transformed any initial state $\rho (0)$ to $\rho (t)$, i.e.,  $\rho (t)={{\phi }_{t}}\left[ \rho (0) \right]$. The map is expected to be completely positive or at least positive to ensure that $\phi_t$ maps physical states to physical states. 
 This preserves the Hermiticity and the trace of operators. The mathematical properties of the map are discussed mainly in the quantum information community. By expanding both $\phi_t\left[ \rho (0) \right]$ and $\rho(0)$ in the basis of the $n^2$ operators $G_k$, the map may be expressed in matrix notation $\rho (t)=\sum\nolimits_{k,l=0}^{{{n}^{2}}-1}{{{F}_{lk}}}(t)Tr\left[ {{G}_{l}}\rho (0) \right]{{G}_{k}}$, i.e., as a function of the $\mathbf{F}(t)$ matrix discussed above and easily computed by HEOM. 
 In order to analyze the positivity, it is rather the $n^2 \times n^2$ Choi matrix \cite{choi1975,Kye2022, Anderson2014} that is used. It corresponds to the expansion of the map in the basis set of the $n^2$ projector matrices $ \lvert j \rangle \langle  k \rvert$ related to a basis of the system Hilbert space in place of the generators $G_n$. The map is completely positive if and only if the eigenvalues of the Choi matrix are positive. 
 This Choi matrix would be easily obtained by HEOM by a procedure similar to that providing the $\mathbf{F}(t)$ matrix via the propagation of the $n^2$ projector matrices $\lvert j \rangle \langle  k \rvert$. To our knowledge, this systematic analysis has not been carried out with HEOM and could be interesting. In our previous works, we have mainly computed with HEOM $\mathbf{F}(t)$ \cite{Chevet2018, Atabek2017} and the canonical rates $\gamma _k(t)$ \cite{MangaudMeier2017,MDL2018}. Anyhow, we did not detect a loss of conservation of the trace of $\rho_S(t)$ even when dynamics is non-Markovian as shown by the volume of accessible states or by the canonical rates. It is well known that numerical instabilities with non-conservation of the norm may occur mainly for long time in the standard formulation of HEOM \cite{Reichman2019} or in the TT approach due to the variational approach \cite{Shi2020} and limitation of the tensor ranks \cite{Borrelli2021}.}
 
 \textcolor{black}{The control of open quantum systems \cite{Koch2016} has aroused a renewed interest mainly in the context of quantum technologies that rely on the coherent manipulation and transfer of information, encoded in quantum states \cite{Maniscalco2013, Mirkin2019, Sugny2022qt}. Non-Markovianity is expected to be a resource to improve the control by exploiting the transitory flow back. The HEOM formalism has proven its efficiency when it is coupled with different control strategies, in particular with optimal control protocols \cite{MDL2018,Jaouadi2022}.}

\section{HEOM in Tensor-Train format}
 In this section, we summarize the main relations of the tensor train formalism (also named matrix product state (MPS)) that is an interesting way for representing a high-dimensional tensor as the one we are using for HEOM. The idea of this TT format is to decompose the tensor into a network of low-dimensional tensors called TT cores coupled in a chain. MPS has received a growing interest in the quantum physics community \cite{Chin2016,SchroderChin2019,Chin2019,Dunnett2021,Dunnett2021bis, Baiardi2019,Ma2019,Plenio2019,Batista2022,Schmidt2022}. The application in HEOM was already suggested by Shi \cite{Shi2018,Shi2020} who also uses the Tucker representation \cite{Shi2021} and later by Borelli \cite{Borrelli2021,Borrelli_Gelin2021}. We present a pedagogical survey showing the way the super-operators involved in the TT-HEOM formalism are written and by expliciting them in a simple case of a two-level system coupled to a bath with a spectral density fitted by an Ohmic Lorentzian (Eq.(\ref{eq:Johmic})) leading to two artificial decay modes. \textcolor{black}{The Appendix gathers the main steps for the encoding in python.}

\subsection{Representation of the ADOs}
When the system is a $n$-state case, each ADO is a $n \times n$ matrix that may be reshaped in a super-vector with $n^2$ elements $ \bar{\rho}^\alpha_{\textbf{n}}$ where $\alpha \in [1,n^2]$ stands for a \textcolor{black}{($a,b$)} element of the \textcolor{black}{system} density matrix \textcolor{black}{($a,b \in [1,n]$)}. The global index $\mathbf{n}$ \textcolor{black}{corresponds as in}  Eq.(\ref{eq:indicen}) \textcolor{black} to the occupation number in each decay mode. When the TT format is adopted, each occupation number $n_k$ runs from 0 to $n_{max}$. The total number of matrices is then larger than when the hierarchy is truncated at a given level $L$ in the standard formulation. Each element of the high-dimensional array $\mathbf{\bar{\rho }}$ is written in TT-format as :
\begin{align}
 &\bar{\rho}^{\alpha}_{\textbf{n}} \approx \sum_{j_0} \sum_{j_1} \cdots \sum_{j_k} \cdots \sum_{j_{K+1}} A_0(j_0, \alpha, j_1)
 \nonumber \\
 &\times A_1 (j_1, n_1, j_2) \cdots A_k (j_{k}, n_k, j_{k+1})
\nonumber \\
&\times \cdots A_K (j_{K}, n_K, j_{K+1}).
\label{eq:rhobar}
\end{align} 
The summation index $j_k$ goes from 1 to $r_k$ where $r_k$ is the rank also called the bond ($r_0 = r_{K+1} = 1$ for dimensionality consistency). $K$ is the number of decay modes. $A_k$ are the cores, i.e. arrays of dimension $r_{k} \times n_{dim} \times r_{k+1}$ where $n_{dim} = n^2$ for $k=0$ and $n_{dim} = n_{HEOM}=n_{max}+1$ for $k \neq 0$ ($n_{HEOM} = L + 1$ where $L$ is the hierarchy level). The choice of the rank $r_k$ is crucial and convergence must be carefully checked. Complex baths (and realistic ones) often need many decay modes \textcolor{black}{and high hierarchy level}: $K$ and $L$ increase dramatically which leads to heavy simulations. 

\textcolor{black}{TT} decomposition allows a priori to deal with this dimensionality curse. Indeed, instead of using an array with $n^2 \times {n_{HEOM}}^K$ in a standard approach, the tensor train decomposition stores $n^2 r_1 \times n_{HEOM}  \prod_{k=1}^{K} r_k r_{k+1} $ tensor elements. For instance, if the ranks for every core are all the same ($\forall k \in [1,K] , r_k = r $), the number of elements is $ r (n^2 + n_{HEOM}) + K r^2 n_{HEOM} $ in the TT approach. With the TT formalism, the increase with the number of modes is linear with $K$. This might allow to deal with large values of $K$ and thus extend the capabilities of the HEOM method to describe more complex environments. However, the TT decomposition approximates the real tensor and is exact only if the ranks $r_k$ grow up to infinity. In practise, they are truncated to a sufficiently high value to ensure the computation convergence. Choosing the best ranks is not a trivial task and goes beyond the scope of this article. We discuss this point in section \ref{sec:dynTT}.

\begin{figure*}
\centering
\includegraphics[width =0.8\textwidth]{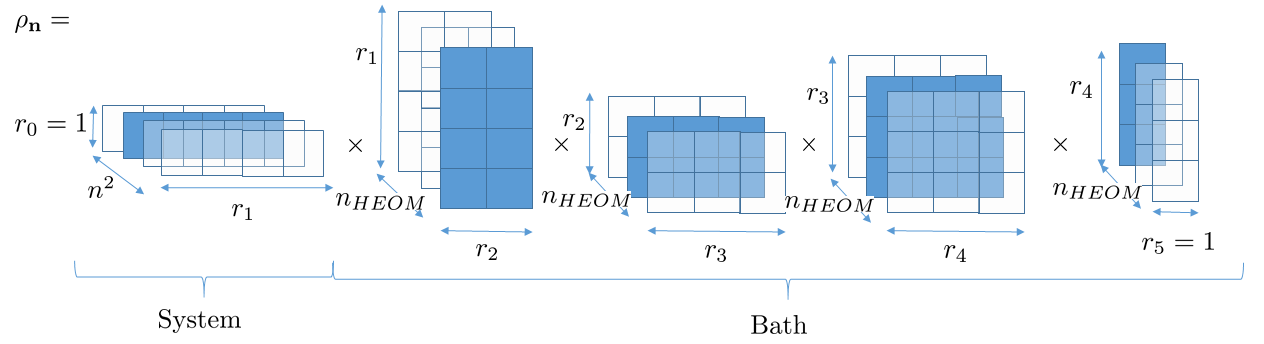}
\includegraphics[width =0.65\textwidth]{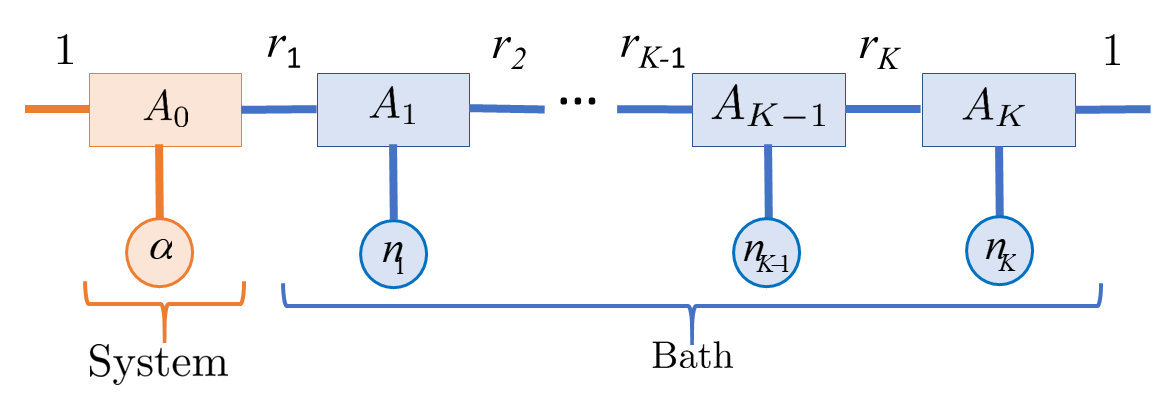}
\caption{Upper panel : Schematic representation of a tensor train with the HEOM formalism for a two-level system ($n=2$) with a bath  modelled by two single Tannor-Meier Lorentzian functions with no Matsubara frequencies ($K = 4$) at a hierarchy level of $n_{HEOM} = 3$. Individual matrices or vectors are called cores of the tensor (and often denoted as $A_k$). Arbitrary values have been chosen here for the tensor ranks: $r_1 =4$, $r_2 = 2$, $r_3=3$ and $r_4 = 4$.  For instance, the tensor element $\rho_{2,3,2,2,1}$, which corresponds to $\rho^{12}_{3,2,2,1}$, i.e. the element 1,2 of the ADO for occupation number $\mathbf{n}=3,2,2,1$, is computed by performing vector-matrix, matrix-matrix or matrix-vector products with the blue cores on the figure. As the initial and final tensor ranks ($ r_0 = r_{K+1} = 1$) are equal to one, the final result is a scalar  \textcolor{black}{number}.
Lower panel : Tensor train representation of the ADOs. $A_k$ are the cores of the tensor. The circles represent the physical legs. $\alpha$ runs from 1 to $n^2$ where $n$ is the number of states in the system and  $n_k$ is the index (occupation number) for each decay mode that runs from 0 to $n_{max}$. The rectangles are matrices $r_{k-1} \times r_k$ where the rank $r_k$ is also called the bond. }
\label{fig:tensortrain}
\end{figure*}

Tensor representations are schematized in figure \ref{fig:tensortrain}.

\subsection{HEOM super-Liouvillian }
\label{section:TT-HEOM}
The HEOM equations (\ref{eq:HEOMshort}) and (\ref{eq:HEOMshortNb})
contain a Liouvillian operator related to the Hamitonian of the system, a damping term, a term coupling to matrices of a higher level in the hierarchy and a term coupling to a lower level. In super-operator notation, one has:
\begin{equation}
\mathcal{L} = \mathcal{L}_S + \sum_{k'=1}^{K} \left( \mathcal{L}_{k'} + \mathcal{L}_{k'+} + \mathcal{L}_{k'-} \right) 
\label{eq:lioutt}
\end{equation} 
where, $k'=(k,b)$ is a collective index which adresses both the index of the correlation function terms ($k \in \left[1, n_{cor,b} \right]$) and the bath $b$. 

For pedagogical purpose, in this section, we give the expression of the different contributions to the super-operator of the forward propagation with TT for a simple example. We consider a two-level system ($n = 2$). \textcolor{black}{The excited state is coupled to a single bath that tunes the energies and is coupled diagonally to the system. The spectral density is a two-pole Lorentzian leading to two decay modes ($K=2$). We assume a high temperature so there is no Matsubara term.} The corresponding coupling operator is 
\begin{equation}
{{S}_{k}}=\left( \begin{matrix}
   0 & 0  \\
   0 & 1  \\
\end{matrix} \right)
\label{equation:operS}
\end{equation}
for \textcolor{black}{decay mode} $k=1,2$. HEOM is treated at order two  $n_{HEOM}=2$, i.e., for each mode, the occupation number $n_k$ may take the value 0 or 1 ($n_{max} = 1$).  The hierarchy contains 4 \textcolor{black}{$(2 \times 2)$} matrices related to occupation numbers 00, 01, 10 and 11. The super-operator is \textcolor{black}{then a $(16 \times 16)$} matrix in \textcolor{black}{this example}.

All the elements of all the ADOs (\textcolor{black}{$n \times n$} matrices) are reshaped in a supervector that contains $n^2$ groups corresponding to a given element of the system density matrix \textcolor{black}{$\alpha = (a,b)$}  when the ADO indexes of $\left\{ {{n}_{1}},..,{{n}_{k}},.,{{n}_{K}} \right\}$ run from $0$ to $n_{max}$ beginning by the last one. In the example with a two-level system, two decay modes ($K=2$) and $n_{max} = 1$, there are 4 groups labelled 00, 01, 10 and 11 (see figure \ref{fig:matL}.

\subsubsection{Matrix $\mathcal{L}_{S}$}
\begin{figure}
 \centering
\includegraphics[width =1.\columnwidth,height=6.5cm]{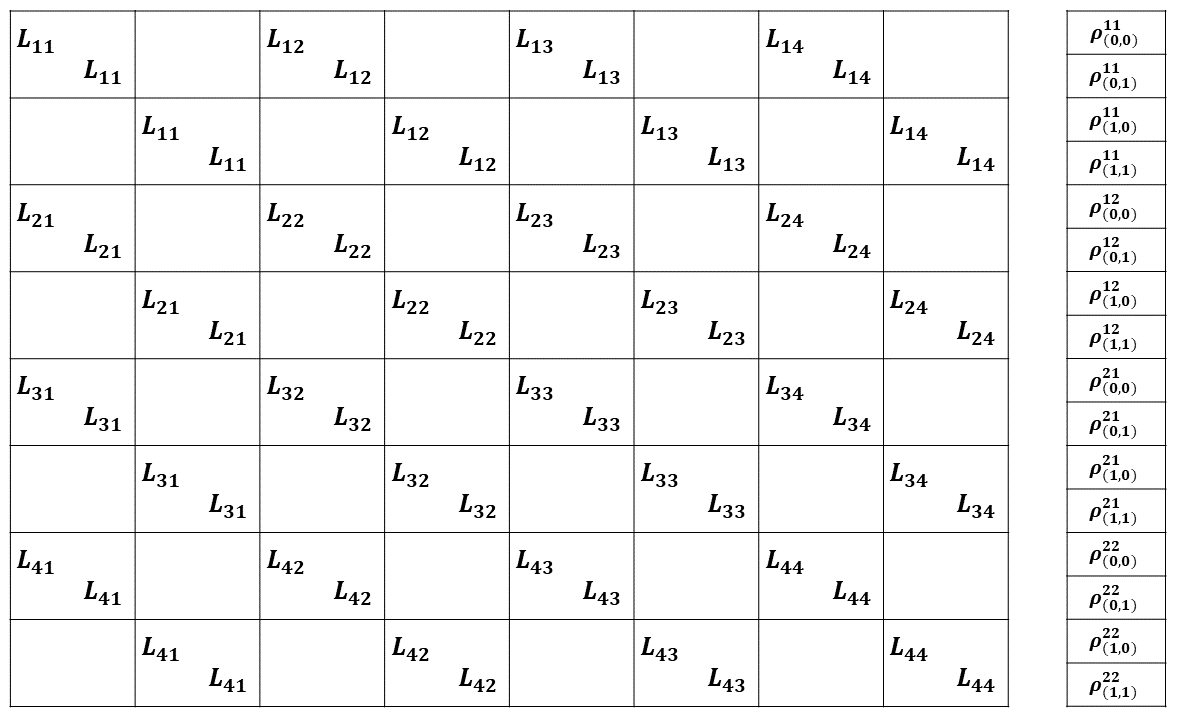}
\caption{Matrix $\mathcal{L}_S$ (Eq.(\ref{eq:LS})). The $L_{ij}$ elements are the concise notation for $(\mathcal{L}_{S(ADO)})_{ij}$ defined in Eq.(\ref{eq:liouSADO}). The empty regions correspond to zero.   }
\label{fig:matL}
\end{figure}

 The expression of the $n^2 \times n^2$ super-operator for the system Liouvillian, which must act on the system density matrix and on each auxiliary density operator (ADO) is straightforward ($\otimes$ denotes here the Kronecker product):
\begin{align}
  & {\mathcal{L}_{S(ADO)}}=-i(H\otimes {{I}_{n}}-{{I}_{n}}\otimes H) \nonumber \\ 
 & =-i\left( \begin{matrix}
   0 & -{{H}_{12}} & {{H}_{12}} & 0  \\
   -{{H}_{21}} & {{H}_{11}}-{{H}_{22}} & 0 & {{H}_{12}}  \\
   {{H}_{21}} & 0 & {{H}_{22}}-{{H}_{11}} & -{{H}_{12}}  \\
   0 & {{H}_{21}} & -{{H}_{21}} & 0  \\
\end{matrix} \right).
\label{eq:liouSADO}  
\end{align}
The expression of the superoperator $\mathcal{L}_{S}$ is then
\begin{equation}
\mathcal{L}_S = \mathcal{L}_{S(ADO)} \otimes \prod_{k''=1}^K I_{n_{HEOM}} 
\label{eq:LS}
\end{equation}
becoming $\mathcal{L}_S = \mathcal{L}_{S(ADO)} \otimes I_2 \otimes I_2$ in the example with $K = 2$ and $n_{HEOM}=2$. The corresponding matrix is given in figure \ref{fig:matL} where we have adopted the concise notation $L_{ij}=(\mathcal{L}_{S(ADO)})_{ij}$.

\subsubsection{Damping Liouvillian $\mathcal{L}_{damp}$}

\begin{figure*}
\centering
\begin{tabular}{cc}
\includegraphics[width =1.\columnwidth,height=6cm]{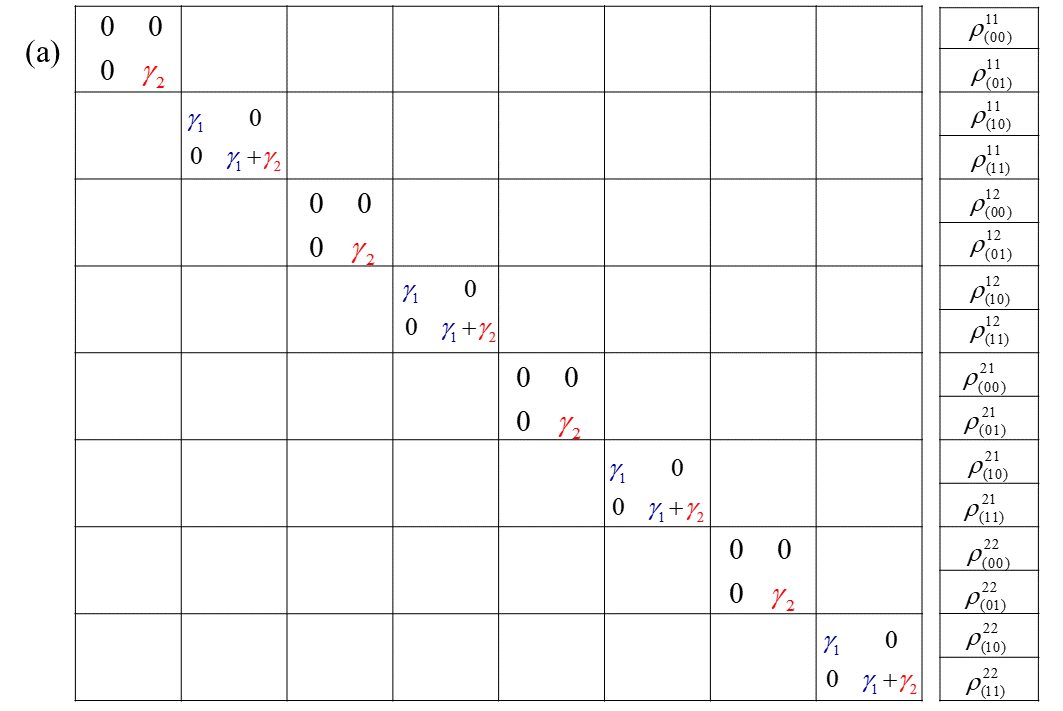}  &
\includegraphics[width =1.\columnwidth,height=6cm]{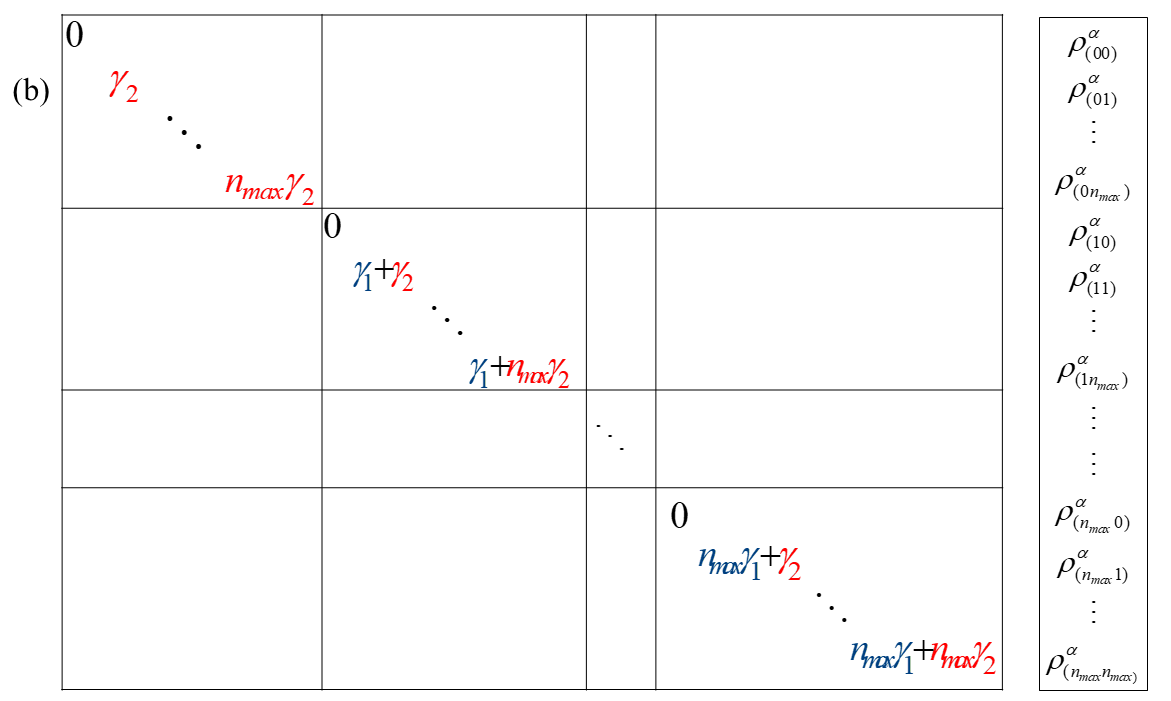}  \\ 
\end{tabular} 
\caption{Color on line. Damping superoperator $(\mathcal{L}_1+\mathcal{L}_2)/i$ (Eqs.(\ref{align:matLdamp1}) and (\ref{align:matLdamp2})). (a) Example of a two-state system with two artificial decay modes and $n_{max}=1$. $\mathcal{L}_1$ in blue and $\mathcal{L}_2$ in red. (b) Block related to a particular element $\rho^{\alpha}_{\mathbf{n}}$ where $\alpha = ij$, when $n_{max}>1$. All the blocks $(ij)$ have the same structure}
\label{fig:matL1L2}
\end{figure*}

The damping term is $\mathcal{L}_{damp}=\sum_{k'=1}^{K} \left( \mathcal{L}_{k'} \right)$ where
\begin{equation}
\mathcal{L}_{k'} =  i \gamma_{k'} I_{n^2} \otimes \prod_{k''=1}^K  M_{k''}  
\end{equation} 
with $M_{k''}  = I_{n_{HEOM}}$ if $k'' \neq k'$ and $M_{k'',lm}  = (l-1)$ $\delta_{l,m}$ if $k''=k'$ ($l,m \in \left[ 1,n_{HEOM} \right]$). 
It is diagonal in the superoperator representation. For each decay mode $k'$, all the elements of the matrices ${{\rho }_{\mathbf{n}}}$ are multiplied by the decay rate ${{\gamma }_{k'}}$ times the occupation number ${{n}_{k'}}$. In the two-level and two-decay case with $n_{HEOM}=n_{max}$, one has:
\begin{align}
  & {\mathcal{L}_{1}}=i{{\gamma }_{1}}{{I}_{{{n}^{2}}}}\otimes M\otimes {{I}_{{{n}_{NHEOM}}}} \nonumber \\ 
 & =i{{\gamma }_{1}}\left( \begin{matrix}
   1 & {} & {} & {}  \\
   {} & 1 & {} & {}  \\
   {} & {} & 1 & {}  \\
   {} & {} & {} & 1  \\
\end{matrix} \right)\otimes \left( \begin{matrix}
   0 & {} & {} & {}  \\
   {} & 1 & {} & {}  \\
   {} & {} & \ddots  & {}  \\
   {} & {} & {} & {{n}_{max}}  \\
\end{matrix} \right) \nonumber \\
&\otimes \left( \begin{matrix}
   1 & {} & {} & {}  \\
   {} & 1 & {} & {}  \\
   {} & {} & \ddots  & {}  \\
   {} & {} & {} & 1  \\
\end{matrix} \right)
\label{align:matLdamp1}
\end{align}
and
\begin{align}
  & {\mathcal{L}_{2}}=i{{\gamma }_{2}}{{I}_{{{n}^{2}}}}\otimes {{I}_{{{n}_{NHEOM}}}}\otimes M \nonumber \\ 
 & =i{{\gamma }_{2}}\left( \begin{matrix}
   1 & {} & {} & {}  \\
   {} & 1 & {} & {}  \\
   {} & {} & 1 & {}  \\
   {} & {} & {} & 1  \\
\end{matrix} \right)\otimes \left( \begin{matrix}
   1 & {} & {} & {}  \\
   {} & 1 & {} & {}  \\
   {} & {} & \ddots  & {}  \\
   {} & {} & {} & 1  \\
\end{matrix} \right) \nonumber \\
&\otimes \left( \begin{matrix}
   0 & {} & {} & {}  \\
   {} & 1 & {} & {}  \\
   {} & {} & \ddots  & {}  \\
   {} & {} & {} & {{n}_{max}}  \\
\end{matrix} \right)  
\label{align:matLdamp2}
\end{align}
When $n_{max} = 1$ and thus $n_{NHEOM} = 2$, ${{M}}=\left( \begin{matrix}
   0 & {}  \\
   {} & 1  \\
\end{matrix} \right)$ and  ${{I}_{{{n}_{NHEOM}}}} = I_2$. The corresponding $\mathcal{L}_{damp}$ matrix divided by $i$ is given in figure \ref{fig:matL1L2}. 

\subsubsection{Matrix $\mathcal{L}_+$} 
\begin{figure}
 \centering
\includegraphics[width =1.\columnwidth,height=6cm]{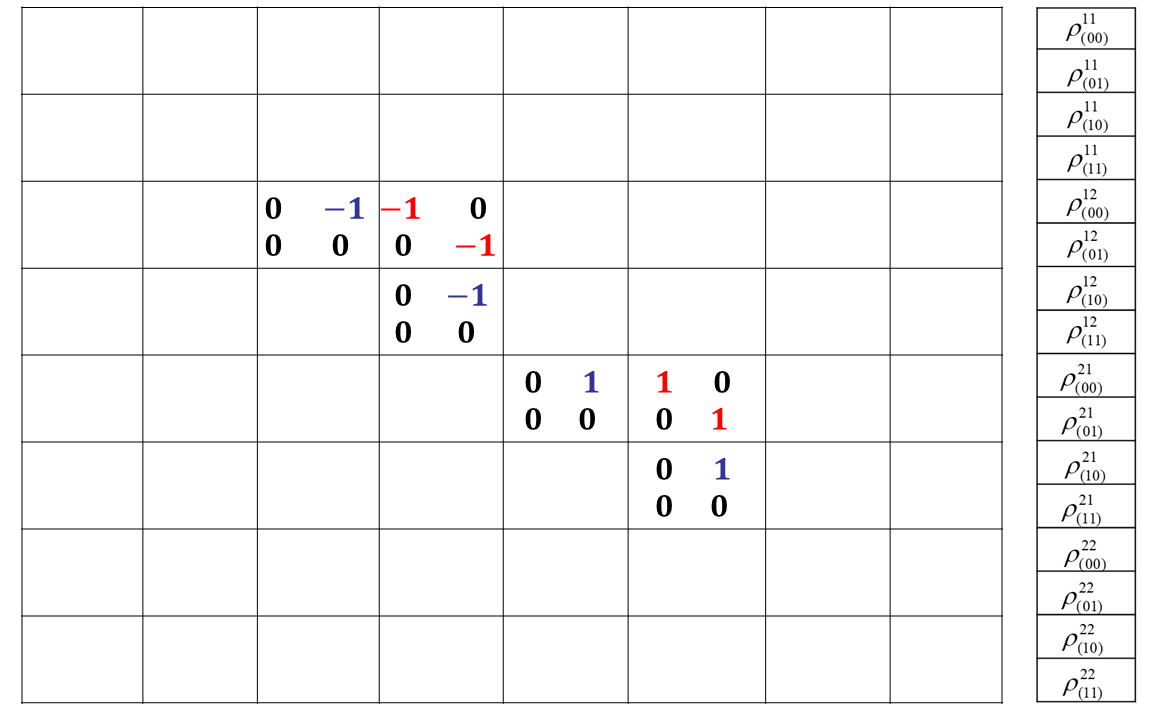}
\caption{Color online. Matrix $(\mathcal{L}_{1+}+\mathcal{L}_{2+})/(-i)$ in the case of a two-state system with two artificial decay modes \textcolor{black}{and} $n_{max}=1$. $\mathcal{L}_{1+}$ in blue and $\mathcal{L}_{2+}$ in red. }
\label{fig:matL+}
\end{figure}

Each term $\mathcal{L}_{k'+}$ addressing the upper level of the hierarchy in Eqs.(\ref{eq:HEOMshort}) or (\ref{eq:HEOMshortNb}) corresponds to the superoperator
\begin{equation}
\mathcal{L}_{k'+} = -i( S_{k'} \otimes I_{n} -  I_{n} \otimes S_{k'} )  \otimes \prod_{k''=1}^K  M'_{k''}.
\label{eq:liou+}
\end{equation}
where $M'_{k''}  = I_{n_{HEOM}}$ if $k'' \neq k'$ and $M'_{k'',lm}  =  \delta_{l+1,m} $ if $k''=k'$ ($l,m \in \left[ 1,n_{HEOM} \right]$). 
We consider the case with a single bath with a system-bath coupling operator $S$ (Eq.(\ref{equation:operS})). Then $S_{k'}$ is independant of $k'$. The factor $Q_+=( S \otimes I_{n} -  I_{n} \otimes S )$ in the two-state case with $n = 2$ is:
\begin{align}
Q_+=\left( \begin{matrix}
   0 & 0 & 0 & 0  \\
   0 & -1 & 0 & 0  \\
   0 & 0 & 1 & 0  \\
   0 & 0 & 0 & 0  \\
\end{matrix} \right).
\end{align}
The two contributions $\mathcal{L}_{k'+}$ for two decay modes and \textcolor{black}{$n_{max} > 1$} are:
\begin{align}
  & {\mathcal{L}_{1+}}=-iQ_+\otimes M'\otimes {{I}_{{{n}_{NHEOM}}}} \nonumber \\ 
 & =-i\left( \begin{matrix}
   0 & 0 & 0 & 0  \\
   0 & -1 & 0 & 0  \\
   0 & 0 & 1 & 0  \\
   0 & 0 & 0 & 0  \\
\end{matrix} \right)\otimes \left( \begin{matrix}
   0 & 1 & {} & {}  \\
   {} & 0 & \ddots  & {}  \\
   {} & {} & \ddots  & 1  \\
   {} & {} & {} & 0  \\
\end{matrix} \right) \nonumber \\
&\otimes \left( \begin{matrix}
   1 & {} & {} & {}  \\
   {} & 1 & {} & {}  \\
   {} & {} & \ddots  & {}  \\
   {} & {} & {} & 1  \\
\end{matrix} \right)  
\label{align:L1+}
\end{align}
and
\begin{align}
  & {\mathcal{L}_{2+}}=-iQ_+\otimes {{I}_{{{n}_{NHEOM}}}}\otimes M' \nonumber \\ 
 & =-i\left( \begin{matrix}
   0 & 0 & 0 & 0  \\
   0 & -1 & 0 & 0  \\
   0 & 0 & 1 & 0  \\
   0 & 0 & 0 & 0  \\
\end{matrix} \right)\otimes \left( \begin{matrix}
   1 & {} & {} & {}  \\
   {} & 1 & {} & {}  \\
   {} & {} & \ddots  & {}  \\
   {} & {} & {} & 1  \\
\end{matrix} \right) \nonumber \\
&\otimes \left( \begin{matrix}
   0 & 1 & {} & {}  \\
   {} & 0 & \ddots  & {}  \\
   {} & {} & \ddots  & 1  \\
   {} & {} & {} & 0  \\
\end{matrix} \right).
 \label{align:L2+} 
\end{align}
When $n_{max}=1$, ${{I}_{{{n}_{NHEOM}}}} = I_2$ and $M'=\left( \begin{matrix}
   0 & 1  \\
   0 & 0  \\
\end{matrix} \right)$
The corresponding matrix of the sum $(\mathcal{L}_{1+}+\mathcal{L}_{2+})/(-i)$ is displayed in figure \ref{fig:matL+}. By comparing with Eq.(\ref{align:HEOM}), the contribution to ${{\dot{\rho }}_{00}}$ of the term related to the upper ADOs with only single excitation is $-i\left[ S,{{\rho }_{10}}+{{\rho }_{01}} \right]$, i.e., $\dot{\rho }_{00}^{12}\to -\rho _{01}^{12}-\rho _{10}^{12}$, $\dot{\rho }_{00}^{21} \to \rho _{01}^{21}+\rho _{10}^{21}$ and $\dot{\rho }_{00}^{11}=\dot{\rho }_{00}^{22} \to 0$. \textcolor{black}{Matrix-vector products of lines 5 and 9 with the column vector $\rho^\alpha_{\textbf{n}}$} in figure \ref{fig:matL+} provide these expressions for the contribution to $\dot{\rho }_{00}^{12}$ and $\dot{\rho }_{00}^{21}$ respectively. The results for $\dot{\rho }_{00}^{11}$ and $\dot{\rho }_{00}^{22}$ \textcolor{black}{can be obtained in the same way} at lines 1 and 13.

\subsubsection{Matrix $\mathcal{L}_-$} 
\begin{figure}
 \centering
\includegraphics[width =1.\columnwidth,height=6cm]{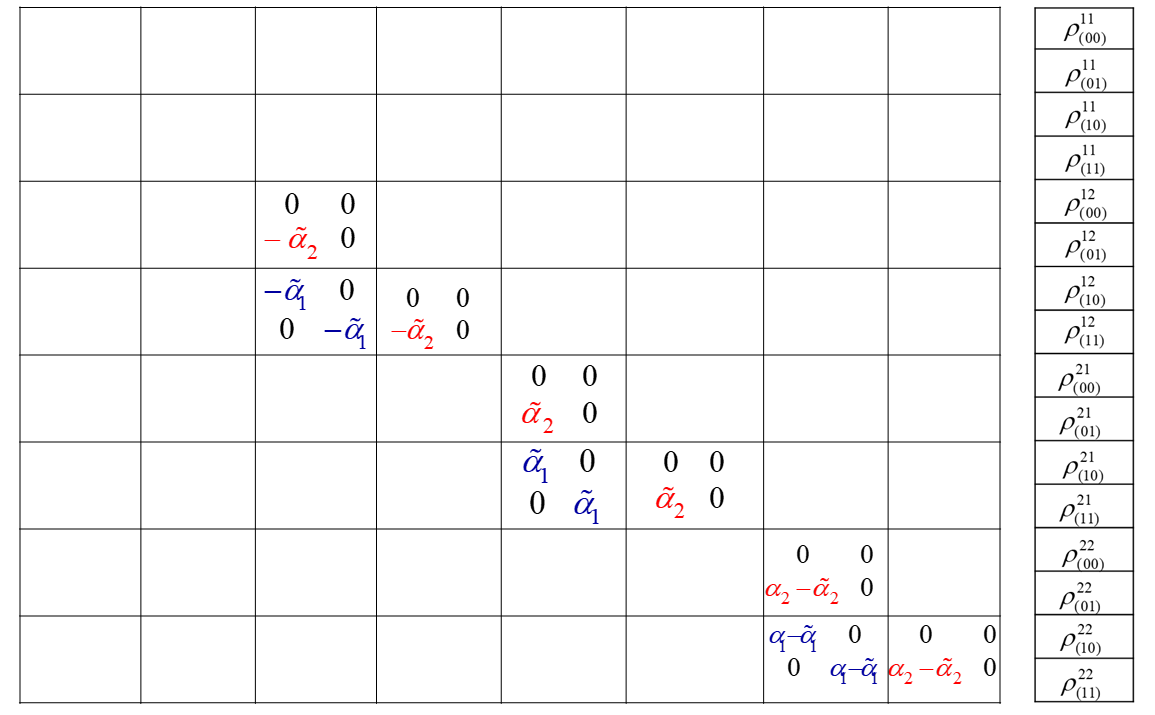}
\caption{Color on line. Matrix $(\mathcal{L}_{1-}+\mathcal{L}_{2-})/(-i)$ in the case of a two-state system with two artificial decay modes \textcolor{black}{and} $n_{max}=1$. $\mathcal{L}_{1+}$ in blue and $\mathcal{L}_{2+}$ in red.  }
\label{fig:matL-}
\end{figure}

The superoperator connecting the ADOs with a lower layer in the hierarchy involves the operator:
\begin{align}
\mathcal{L}_{k'-} = & -i(\alpha_{k',t/c} S_{k'} \otimes I_{n} -\tilde \alpha_{k',t/c} I_{n} \otimes S_{k'} )
\nonumber \\
&\otimes \prod_{k''=1}^K  M''_{k''}
\end{align}
where $M''_{k''}  = I_{n_{HEOM}}$ if $k'' \neq k'$ and $M''_{k'',lm} = (l-1)$ $\delta_{l-1,m}$ if $k''=k'$  ($l,m \in \left[ 1,n_{HEOM} \right]$).
In the two-level case with $S_{k'}$ independent of $k'$, for each mode, one has the factor:
\begin{align}
  & {{Q}_{-k'}}={{\alpha }_{k'}}{{S}}\otimes {{I}_{n}}-{{{\tilde{\alpha }}}_{k'}}{{I}_{n}}\otimes {{S}} \nonumber \\ 
 & ={{\alpha }_{k'}}\left( \begin{matrix}
   0 & 0  \\
   0 & 1  \\
\end{matrix} \right)\otimes \left( \begin{matrix}
   1 & 0  \\
   0 & 1  \\
\end{matrix} \right)-{{{\tilde{\alpha }}}_{k'}}\left( \begin{matrix}
   1 & 0  \\
   0 & 1  \\
\end{matrix} \right)\otimes \left( \begin{matrix}
   0 & 0  \\
   0 & 1  \\
\end{matrix} \right) \nonumber \\ 
 & =\left( \begin{matrix}
   0 & 0 & 0 & 0  \\
   0 & {{{- \tilde{\alpha }}}_{k'}} & 0 & 0  \\
   0 & 0 & {{\alpha }_{k'}} & 0  \\
   0 & 0 & 0 & {{\alpha }_{k'}}-{{{\tilde{\alpha }}}_{k'}}  \\
\end{matrix} \right).  
\end{align}
The two contributions to the super-Liouvillian when $n_{max} > 1$ are:
\begin{align}
  & {\mathcal{L}_{1-}}={{Q}_{-1}}\otimes M''\otimes {{I}_{{{n}_{HEOM}}}} \nonumber \\ 
 & =\left( \begin{matrix}
   0 & 0 & 0 & 0  \\
   0 & {{{-\tilde{\alpha }}}_{1}} & 0 & 0  \\
   0 & 0 & {{\alpha }_{1}} & 0  \\
   0 & 0 & 0 & {{\alpha }_{1}}-{{{\tilde{\alpha }}}_{1}}  \\
\end{matrix} \right)\otimes \left( \begin{matrix}
   0 & {} & {} & {}  & {}\\
   1 & 0 & {} & {} & {} \\
   {} & {2} & \ddots  & {}  & {}\\
   {} & {} & \ddots  & \ddots  & {}\\
   {} & {} & {} & {{n}_{\max }} & 0  \\
\end{matrix} \right)   \nonumber \\
& \otimes \left( \begin{matrix}
   1 & {} & {} & {}  \\
   {} & 1 & {} & {}  \\
   {} & {} & \ddots  & {}  \\
   {} & {} & {} & 1  \\
\end{matrix} \right)  
\end{align}
and
\begin{align}
& {\mathcal{L}_{2-}}={{Q}_{-2}}\otimes {{I}_{{{n}_{HEOM}}}}\otimes M'' \nonumber \\ 
& =\left( \begin{matrix}
   0 & 0 & 0 & 0  \\
   0 & {{{-\tilde{\alpha }}}_{2}} & 0 & 0  \\
   0 & 0 & {{\alpha }_{2}} & 0  \\
   0 & 0 & 0 & {{\alpha }_{2}}-{{{\tilde{\alpha }}}_{2}}  \\
\end{matrix} \right)\otimes \left( \begin{matrix}
   1 & {} & {} & {}  \\
   {} & 1 & {} & {}  \\
   {} & {} & \ddots  & {}  \\
   {} & {} & {} & 1  \\
\end{matrix} \right)  \nonumber \\
& \otimes \left( \begin{matrix}
   0 & {} & {} & {}  & {}\\
   1 & 0 & {} & {} & {} \\
   {} & {2} & \ddots  & {}  & {}\\
   {} & {} & \ddots  & \ddots  & {}\\
   {} & {} & {} & {{n}_{\max }} & 0  \\
\end{matrix} \right) .  
\end{align}
In the example of the two decay modes with $n_{max} = 1$, one has $M''=\left( \begin{matrix}
   0 & 0  \\
   1 & 0  \\
\end{matrix} \right)$. The matrices for $\mathcal{L}_{1-}$ and $\mathcal{L}_{2-}$ are given in figure \ref{fig:matL-}.

\subsection{Dynamics with Tensor-Train format}
\label{sec:dynTT}
Dynamics is driven by solving
\begin{equation}
\mathbf{\dot{\bar{\rho} }}(t)=\mathcal{L}\mathbf{\bar{\rho} }(t)
\end{equation}
where $\bar{\rho }$ is the full TT-converted vector of the elements of all the ADOs and $\mathcal{L}$ is the super-operator described in Sec.\ref{section:TT-HEOM}. We use the projector-splitting KSL scheme \citep{Lubich2014,Lubich2015,Oseledets2016,Batista2022} implemented in the ttpy package (\verb|tt.ksl.ksl|) \cite{ttpy}. The method is based on the dynamical low-rank approximation which is equivalent to the Dirac-Frenkel time-dependent variational principle used in MCTDH. It consists in using an approximate low-rank tensor with fixed ranks instead of getting a solution with a high rank tensor and then truncate it with singular value decomposition (SVD). To comply \textcolor{black}{with} this goal, the derivative of the approximate low-rank tensor is obtained by orthogonally projecting the derivative of the tensor on the tangent space of the approximate low-rank tensor at its current position. Time-integration is then obtained by a splitting scheme (second order in this work) of the projector (see Refs. \citep{Lubich2014,Lubich2015,Oseledets2016,Batista2022} for more details). An adaptative rank may be necessary during the propagation as proposed in Refs.\cite{DunnettChin2021,Dolgov19}. We have adopted a mixed strategy. The standard Runge-Kutta integrator (written with TT algebra available with the ttpy package) is run after some time steps to allow the increase of the ranks during the propagation. In the application, we use a Runge-Kutta run after 10 timesteps.  

\section{Illustrative Applications }
We give \textcolor{black}{three} examples for which dynamics is driven by the TT method. \textcolor{black}{In the first two cases, the spectral density is continuous and is an Ohmic two-pole Lorentzian function (Eq.(\ref{eq:Johmic})) leading to two artificial bath modes. The third application uses undamped discrete modes. }

\subsection{Population to coherence transfer via a bath}
\begin{figure*}
 \centering
\includegraphics[height=4.5cm]{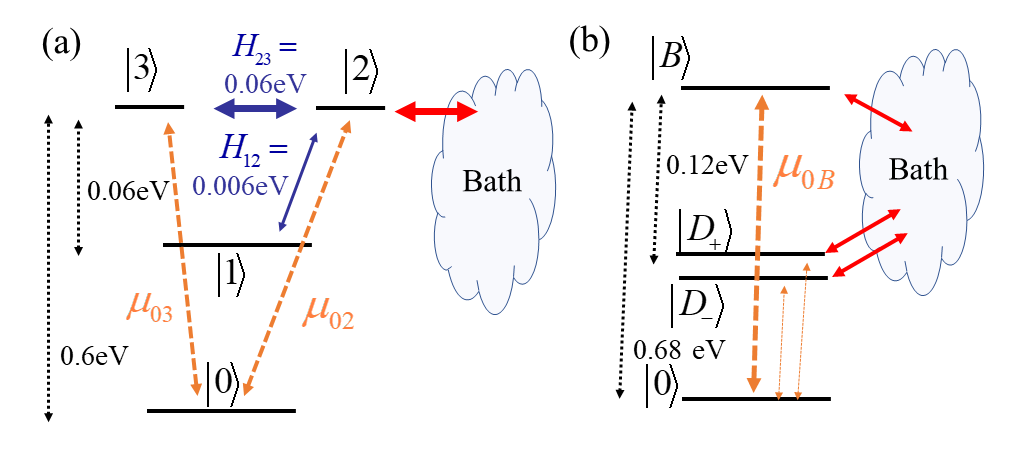}
\caption{Schematic representation of the energy levels, the interstate couplings $H_{ij}$, the dipolar couplings $\mu_{ij}$ and the system-bath couplings of a device in which a long-lasting coherence in a dark doublet may be created by interaction with a bath after excitation of a bright state.  (a) zero-order basis, (b) eigenstates. }
\label{fig:schemecohe}
\end{figure*}

 The zero-order model and the eigenstates are schematized in figure \ref{fig:schemecohe}. This model was introduced to analyze one of the first applications of an experimental ‘quantum simulator’ for molecular quantum dynamics \cite{Potocnik2018}. As we show, using a circuit of qubits to represent a network  of chromophores gives access to strongly non-Markovian regimes of open dynamics, including situations in which strong dissipation actually induces coherent dynamics. Indeed, one of the original motivations for the experiment and analysis of Ref.\citep{ChinPRA2018} was to explore the existence, robustness and uses of coherent transport in photosynthetic light-harvesting proteins, which are described by \textcolor{black}{analogous} models\textcolor{black}{. However, they} are much harder to control compared to superconducting circuits. The Hamiltonian of the active subsystem is:
 \begin{align}
&H_S(t) = H_s + H_{ren} \nonumber  \\
&- \sum_{n=0}^{3}\sum_{m\neq{n}=0}^{3} \mu_{nm} \vert n \rangle \langle m \vert \mathcal{E}(t)
\end{align}
with 
\begin{equation}
H_s = \sum_{n = 0}^{3} \varepsilon _n \vert n \rangle \langle n \vert  + \sum_{n=0}^{3}\sum_{m\neq{n}=0}^{3} H_{nm} \vert n \rangle \langle m \vert 
\end{equation}
 where $H_{ren}$ is the renormalization term given below and $\mu_{ij}$ is the dipolar coupling. The ground state is coupled to the excited states only radiatively, i.e. $H_{0j}=0$ for $j\ne 0$ and only $\mu_{02}$ and $\mu_{03}$ induce the radiative coupling. Two degenerate excited bright states ($\left\vert 2 \right\rangle $ and $\left\vert 3 \right\rangle $) strongly interact by interstate coupling $H_{23}$ and state $\left\vert 2 \right\rangle$ is weakly coupled to a dark lower state $\left\vert 1 \right\rangle $ and to a tuning bath \textcolor{black}{that makes fluctuate the energy}. The strong $H_{23}$ coupling leads to eigenstates that are mainly the bright in phase and the dark out of phase superpositions denoted $\left\vert B \right\rangle $ and $\left\vert D_{+} \right\rangle $ respectively.  Both eigenstates $\left\vert D_{+} \right\rangle $ and $\left\vert D_{-} \right\rangle $ form a dark doublet. Their dipole transition moments $\mu_{0D_{\pm}}$ are very weak, being two orders of magnitude smaller than the transition moment $\mu_{0B}$ to the $\left\vert B \right\rangle $ state. Spontaneous radiative decay is not taken into account in the simulation. 
 
\begin{figure}
 \centering
\includegraphics[width =1.\columnwidth]{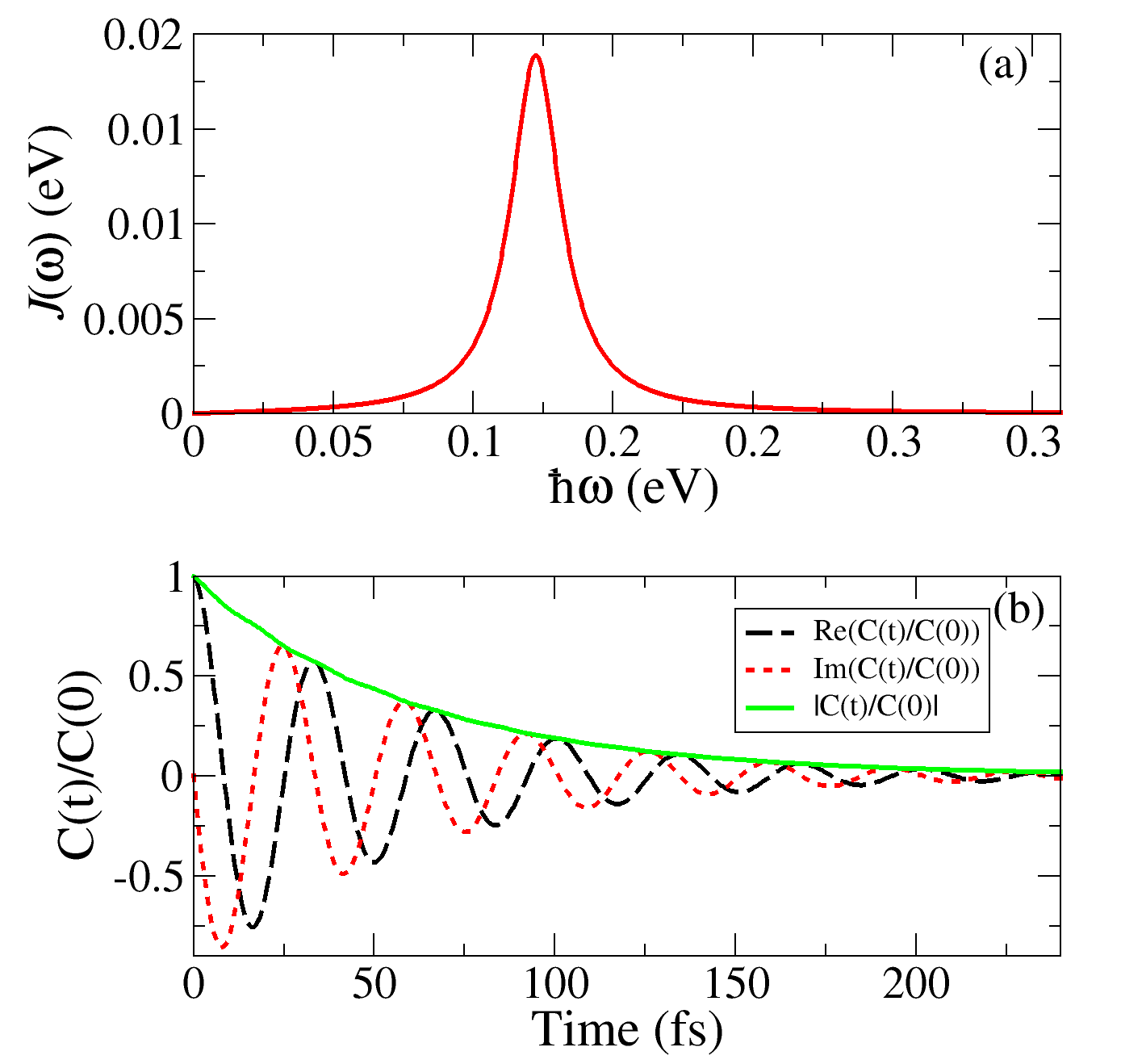}
\caption{(a) Spectral density of the model presented in figure \ref{fig:schemecohe} to illustrate the population-to-coherence transfer. The spectral density maximum is close to the mean $BD$ energy gap; (b) corresponding normalized correlation function $C(t)/C(0)$ (Eq.(\ref{eq:C(t)})) \textcolor{black}{at $T$ = 298K} with real part (dashed line), imaginary part (dots) and modulus (solid line). }
\label{fig:speccorcohe}
\end{figure}

The bath is coupled to state $\left\vert 2 \right\rangle $ only. The corresponding system-bath operator takes the form:
\begin{align}
S=\left( \begin{matrix}
   0 & 0 & 0 & 0  \\
   0 & 0 & 0 & 0  \\
   0 & 0 & 1 & 0  \\
   0 & 0 & 0 & 0  \\
\end{matrix} \right).
\end{align}
 The spectral density is highly structured and peaks nearly at the mean \textcolor{black}{($\Delta_{B-D} = \langle B \vert H_s \vert B \rangle - \langle D \vert H_s \left\vert D \right\rangle$)} transition energy, \textcolor{black}{$\langle D \vert H_s \left\vert D \right\rangle$} being the average energy of states \textcolor{black}{$\langle D_+ \vert H_s \left\vert D_+ \right\rangle$} and \textcolor{black}{$\langle D_- \vert H_s \left\vert D_- \right\rangle$}. It is a single Ohmic Lorentzian (Eq.(\ref{eq:Johmic})) with parameters $p = 2.0 \times 10^{-12} a.u.$, $\Omega = 4.5 \times 10^{-3} a.u.$ and $\Gamma = 4.0 \times 10^{-4}a.u.$. It is represented with the correlation function \textcolor{black}{at $T$ = 298K} in figure \ref{fig:speccorcohe}. The renormalisation energy is $\lambda =(1/\pi )\int_{0}^{\infty }{d\omega J(\omega )/\omega }$. The corresponding renormalization term is $H_{ren} = \lambda \left\vert 2 \right\rangle \left\langle 2 \right\vert$. The system-bath coupling is weak. It is estimated by the ratio \textcolor{black}{$\eta =\lambda /{\Delta_{{(\left\vert B \right\rangle - \left\vert D \right\rangle)}}}$} where \textcolor{black}{$\Delta_{{(\left\vert B \right\rangle - \left\vert D \right\rangle)}}$} is the energy gap corresponding here to the cutoff of the spectral density. It is a perturbative regime, however, the correlation time is long (about 250 fs) due to the peaked shape of the spectral density. It is longer than \textcolor{black}{the Rabi period} of the \textcolor{black}{($\left\vert B \right\rangle - \left\vert D \right\rangle$)} transition that is 34 fs. Dynamics is then non-Markovian \textcolor{black}{with} $n_{HEOM} = 5$.  
In the eigenstate representation (Fig.\ref{fig:schemecohe}(b)), the $S$ operator becomes:
\begin{align}
{{S}_{eigen}}=\left( \begin{matrix}
   0 & 0 & 0 & 0  \\
   0 & 0.26 & 0.25 & 0.36  \\
   0 & 0.25 & 0.24 & 0.34  \\
   0 & 0.36 & 0.34 & 0.50  \\
\end{matrix} \right).
\end{align}

\begin{figure}
\includegraphics[width =1.\columnwidth]{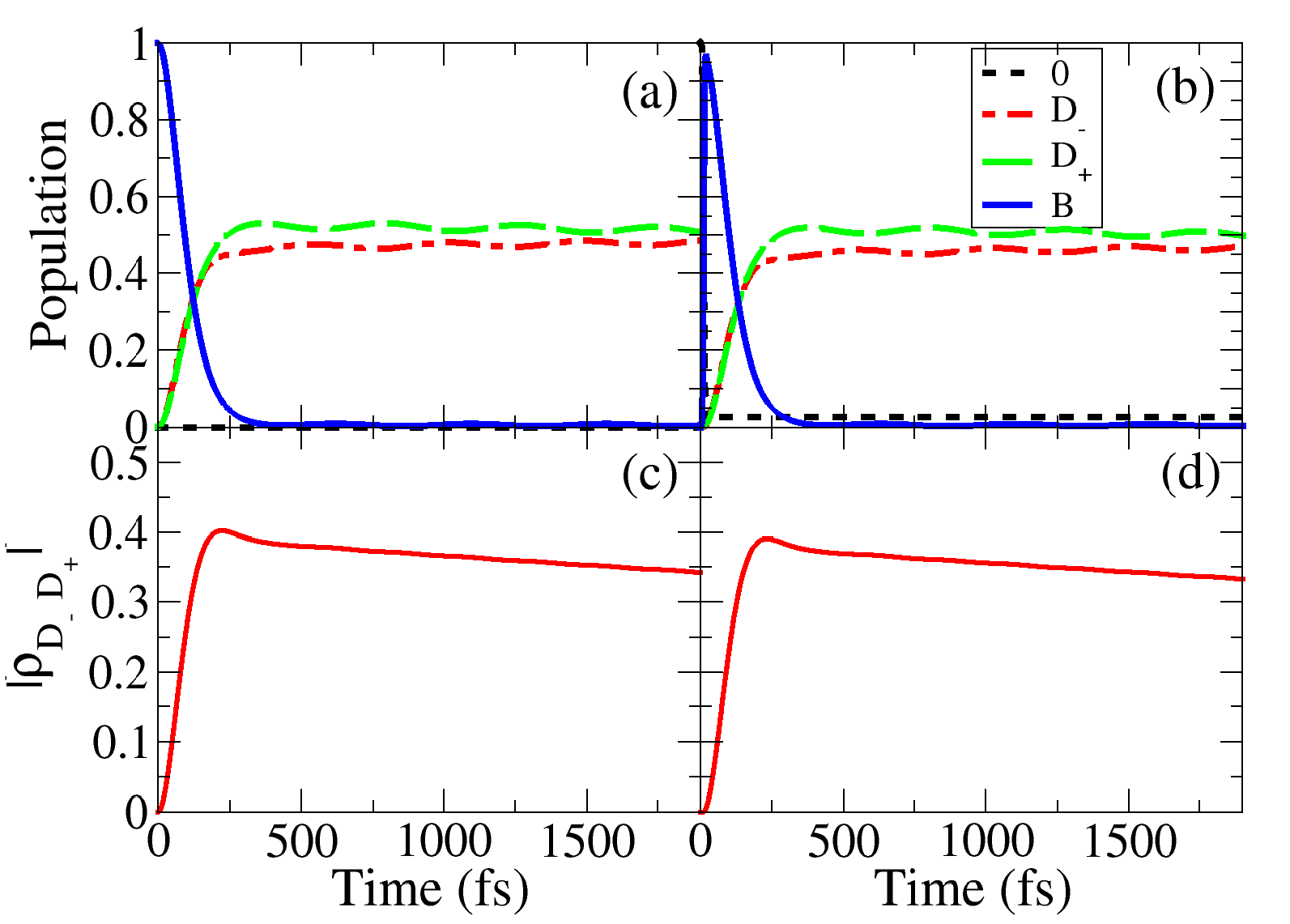}
\caption{Upper panels: populations in the eigenstates $B$, $D_-$ and $D_+$, (a) ideal preparation in the bright state $B$, (b) excitation of the $B$ state from the ground state by a $\pi$ pulse of 24 fs. lower panels: modulus of the coherence $(\rho_{S})_{D_+D_-}$, (c) ideal preparation in the bright state, (d) excitation by a $\pi$ pulse.   }
\label{fig:popcohe}
\end{figure}

In this basis set, the bath is coupled diagonally and off-diagonally to the system. This is an interesting device leading to a population-to-coherence transfer from the bright $\left\vert B \right\rangle $ state to the dark doublet. This process requires exact non-Markovian dynamics taking terms that are neglected in the secular approximation of a Redfield treatment \citep{Lovett2016,ChinPRA2018}. From the analysis in Ref.\citep{ChinPRA2018}, we choose a coupling strength $\eta = 0.013$ relevant to illustrate an efficient population-to-coherence process. The populations in the eigenstates and the modulus of the coherence $(\rho_{S})_{D_+D_-}$ between the two dark states are displayed in figures \ref{fig:popcohe}(a) and \ref{fig:popcohe}(c) when the initial state is the bright state $\left\vert B \right\rangle $. In this ideal case, the population-to-coherence is complete in 250 fs. The coherence is long-lasting and slowly decays in about 25 ps. Figures \ref{fig:popcohe}(b) and \ref{fig:popcohe}(d) compare this ideal preparation with the results obtained by an excitation from the ground state by a laser field of 24 fs. To respect the condition that the area of the oscillating field \textcolor{black}{must be equal to} zero \cite{Brabec2000,Milosevic2006}, the field is then given by $\mathcal{E}(t)=-\frac{\partial \mathcal{A}(t)}{\partial t}$ with the vector potential $\mathcal{A}(t)=\left( \frac{{{\mathcal{E}}_{0}}}{\omega } \right){{\sin }^{2}}\left( \frac{\pi (t-{{t}_{i}})}{\tau} \right)\sin \left( \omega (t-{{t}_{i}})  \right)$, where $t_i$ is the initial time of the pulse, $t_i$ = 0 here, $\tau$ is the pulse duration and $\mathcal{E}_0$ is the field maximum amplitude. The carrier frequency is in resonance with the $\vert 0 \rangle \rightarrow \vert B \rangle $ transition. When the number of cycles is large for a pulse duration longer than about 20 fs, the expression becomes similar to a pulse with a sine square envelope ${\mathcal{E}}(t)={{\mathcal{E}}_{0}}{{\sin }^{2}}\left( \frac{\pi t}{\tau } \right)\cos \left( {{\omega}_{0B}}t \right)$. We use this expression to estimate $\mathcal{E}_0$ providing a $\pi$ pulse for which the integral of the Rabi frequency $\Omega (t) = \mathcal{E}_{0}{{\sin }^{2}}\left( \frac{\pi t}{\tau } \right)\mu_{0B}/\hbar$ is equal to $\pi$ \citep{Thomas1983,Holthaus1994}. $\mu_{0B}$ is the transition dipole. Then ${\mathcal{E}_{0}}=2\pi /({{\mu }_{0B}}{{\tau }})$. This field would induce a complete transfer towards the bright state in the absence of coupling to the bath. A slight modification of the amplitude should be necessary for short pulses with few cycles for which the envelope slighly differs from a sine square.

\begin{figure}
 \centering
\includegraphics[width =1.\columnwidth]{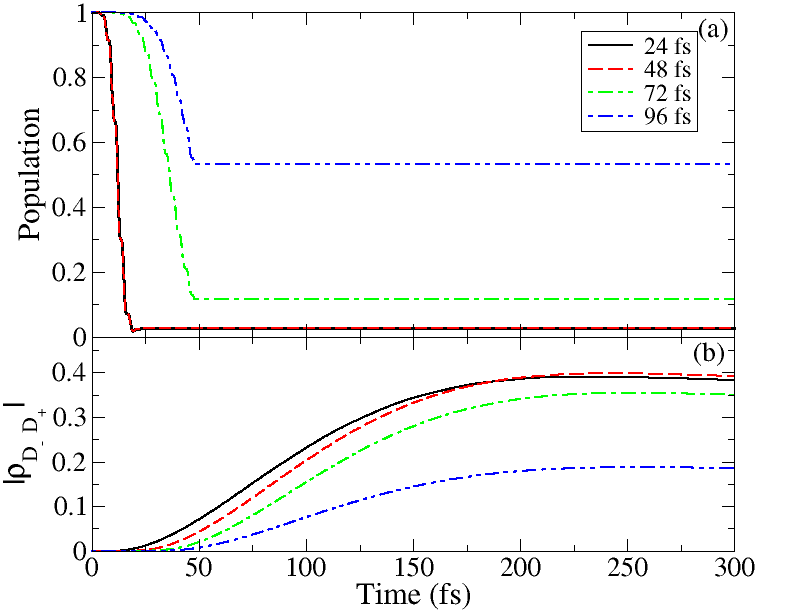}
\caption{Excitation of the bright state $B$ by a $\pi$ pulse of different durations. (a) Population in the ground state, (b) Modulus of the coherence $(\rho_{S})_{D_+D_-}$ generated in the dark doublet . }
\label{fig:coheTmax}
\end{figure}

Figure \ref{fig:coheTmax} shows the influence of the pulse duration $\tau$ on the coherence generation in the dark doublet. The result is close to the ideal case for pulses with $\tau$ in the range 20-50 fs. The decrease of the yield comes from the environment that makes fluctuate the energies.

Simulations are made with a timestep of 0.24 fs and the maximum rank is 10 with $\epsilon = 10^{-15}$. \textcolor{black}{ The temperature is 298 K. No Matsubara term is needed due to the high temperature and the narrow spectral density. No significant change has been observed when carrying out this simulation with 3 additional Matsubara terms. }

\subsection{Simulation of absorption and emission spectra}
HEOM have already been used to simulate stationary absorption \textcolor{black}{$\sigma _{abs}$} and emission \textcolor{black}{$\sigma_{em}$} spectra \cite{Tanimura2006,Tanimura_rev_20,Shi2013,Cao2014,Yan2017,Kramer2018}. They are computed by the linear response theory \textcolor{black}{as:}
\begin{equation}
{{\sigma }_{abs}}(\omega )={\textrm{Re}\int_{0}^{\infty }{dt{{e}^{i\omega t}}T{{r}_{S}}}\left[ \rho _{{{\mu }^{-}}}^{\dagger }(t){{\rho }_{{{\mu }^{-}}}}(0) \right] }
\label{equ:abs}
\end{equation}
\begin{equation}
{{\sigma }_{em}}(\omega )={\textrm{Re}}  \int_{0}^{\infty }{dt{{e}^{i\omega t}}T{{r}_{S}}}\left[ \rho _{{{\mu }^{+}}}^{\dagger }(t){{\rho }_{{{\mu }^{+}}}}({{t}_{eq}}) \right]
\label{equ:em}
\end{equation}
where ${{\mu }^{-}}=\sum\nolimits_{k\ne 0}{{{\mu }_{k}}} \left\vert 0 \right\rangle \left\langle  k \right\vert$ and ${{\mu }^{+}}=\sum\nolimits_{k\ne 0}{{{\mu }_{k}}}\left\vert k \right\rangle \left\langle  0 \right\vert$. The initial conditions for the absorption is ${{\rho }_{{{\mu }^{-}}}}(0)={{\mu }^{-}}{{\rho }_{S}}(0)$ where ${{\rho }_{S}}(0)$ is the reduced density matrix of the system in its ground state and all the ADOs are zero. The stationary emission is \textcolor{black} {computed at} the thermal equilibrium \textcolor{black}{of the emitting state}. The first strategy is to propagate an arbitrary system density matrix with population in the excited states towards the thermally equilibrated state, which is independent of the chosen initial state \citep{Dijkstra2010,Shi2013,Kramer2018}. Then, the initial conditions for the emission is ${{\rho }_{{{\mu }^{+}}}}(t_{eq})={{\mu }^{+}}{{\rho }_{S}}(t_{eq})$ where ${{\rho }_{S}}(t_{eq})$ is the reduced system density matrix at equilibrium and the ADOs take their asymptotic values at time $t_{eq}$. $t_{eq}$ is estimated by verify the population and the average value of the collective bath mode. Another possibility to reach the equilibrium involves propagation with an imaginary time \cite{Pomyalov2010,Tanimura2014,Shi2015}.  
\begin{figure}
 \centering
\includegraphics[width =1.\columnwidth,height=4.5cm]{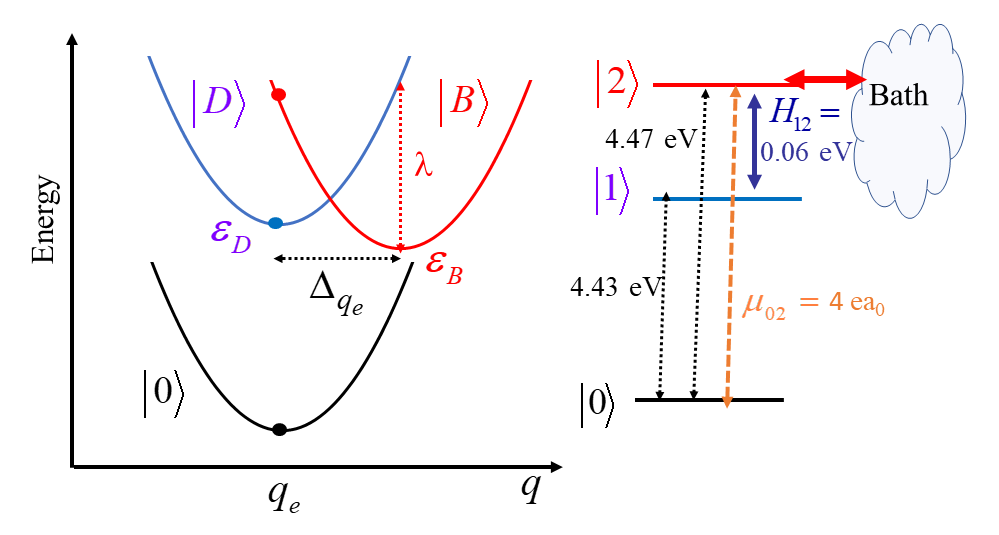}
\caption{ Schematic representation of the diabatic potential energy surface of the excited states used in the simulation of absorption and emission spectra. $q$ is a bath nuclear coordinate and $\Delta_{q_e}$ is the difference of equilibrium positions. The bath is at equilibrium in the ground state and in the dark state $\left\vert D \right\rangle $ since the equilibrium positions are assumed to be the same. The electronic energies of the system correspond to Franck-Condon vertical energies represented by full circles. $\lambda$ is the renormalization energy. Two cases are simulated, the first one with $H_{12} =0$ (the $\left\vert D \right\rangle $ state is spectator) and the second one with ${{H}_{12}}\ne 0$}.
\label{fig:potentiel}
\end{figure}

Figure \ref{fig:potentiel} schematizes the model. According to the value of the interstate coupling $H_{12}$, we consider a single bright state when ${{H}_{12}}=0$ or a two-excited-state case with a nonadiabatic coupling with the dark state when ${{H}_{12}}\ne 0$. The spectra are computed by making the electronic-nuclear partition. The system is composed of two or three electronic states corresponding to vertical energies at the ground state equilibrium geometry. The bath consists in all the nuclear intermolecular and solvent vibrators. Within this partition, the system-bath couplings $c_j=\omega^2_j\Delta_{eq,j}$ depend on the difference ${{\Delta }_{{{q}_{e}}}}$ between the equilibrium position of all the vibrational modes in the ground and excited state. In our example, only state $\left\vert B \right\rangle $ is displaced and therefore coupled to the bath. In the first example ($H_{12} = 0$), the system-bath coupling operator is that given in Eq.(\ref{equation:operS}). In the second case, in the diabatic representation, it becomes
\begin{align}
S=\left( \begin{matrix}
   0 & 0 & 0  \\
   0 & 1 & 0  \\
   0 & 0 & 0  \\
\end{matrix} \right).
\end{align}
We intend to illustrate the strong influence of the spectral density on the Stokes shift by referring to Mukamel’s $\kappa$ parameter \cite{Mukamel1995}
\begin{equation}
\kappa =\frac{\Lambda }{\Delta }.
\label{equ:kappa}
\end{equation}
$\Lambda$ corresponds to the cutoff of the spectral density. ${{\Lambda }^{-1}}$ is an estimation of the bath fluctuation timescale.  ${{\Delta }^{2}}=C(t=0)$ is related to the initial value of the bath correlation function, which is a real value obtained from Eq.(\ref{eq:CtoJ}). $\Delta$ is an estimation of the amplitude of the fluctuations.  

\begin{figure}
 \centering
\includegraphics[width =1.\columnwidth]{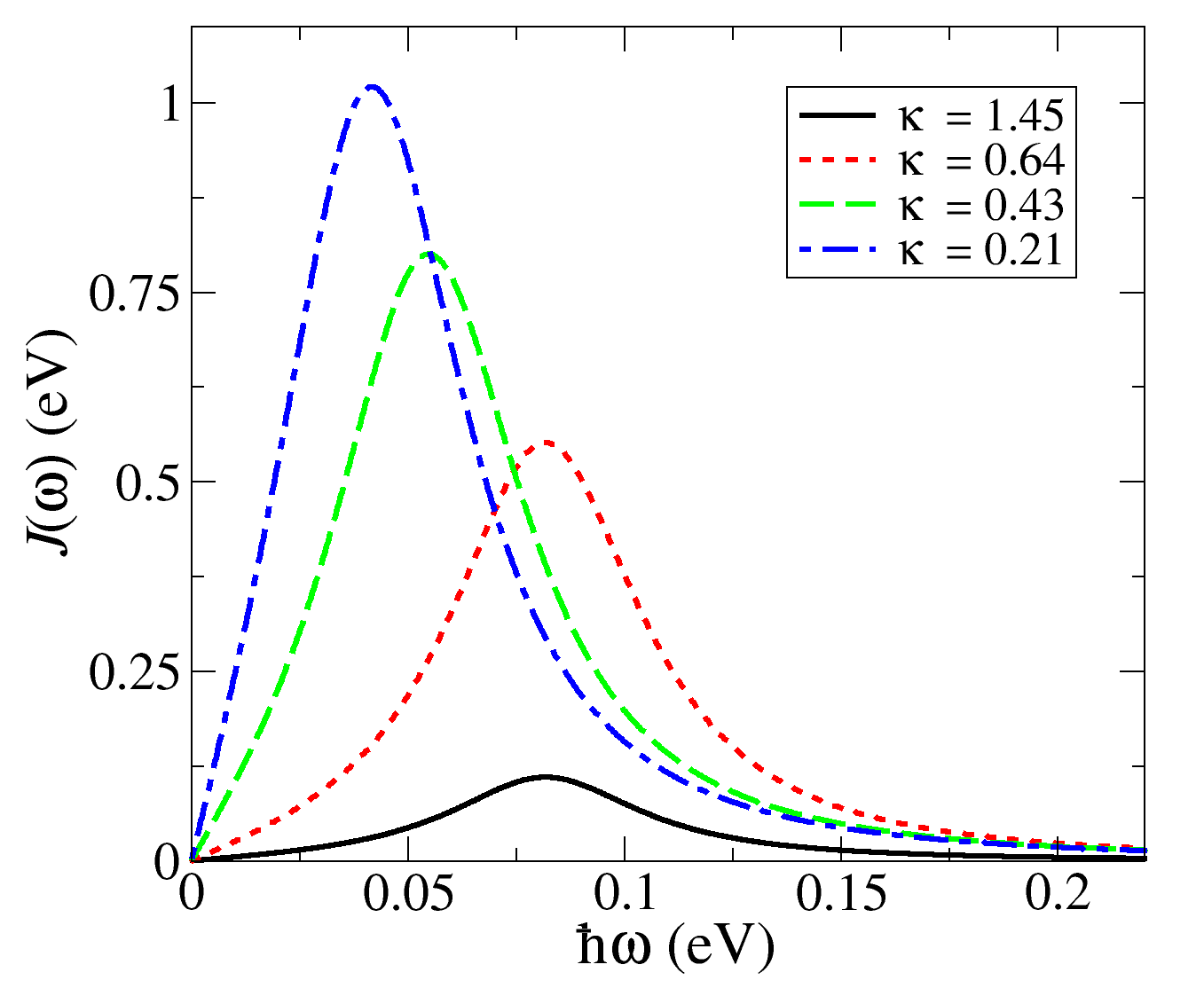}
\caption{Spectral densities used in the simulation of the absorption and emission spectra. $\kappa$ is defined in Eq.(\ref{equ:kappa}). The $\Delta$ parameter is given by the initial value of the correlation function (Eq.(\ref{eq:C(t)})). The cutoff $\Lambda$ is estimated by the energy at the maximum. The renormalization energy $\lambda$ for decreasing values of $\kappa$ are 0.034 eV, 0.17 eV, 0.34 eV and 0.52 eV. }
\label{fig:density}
\end{figure}

Figure \ref{fig:density} displays the four spectral densities adopted in the simulation and the corresponding $\kappa$. 

\begin{figure}
 \centering
\includegraphics[width =1.05\columnwidth]{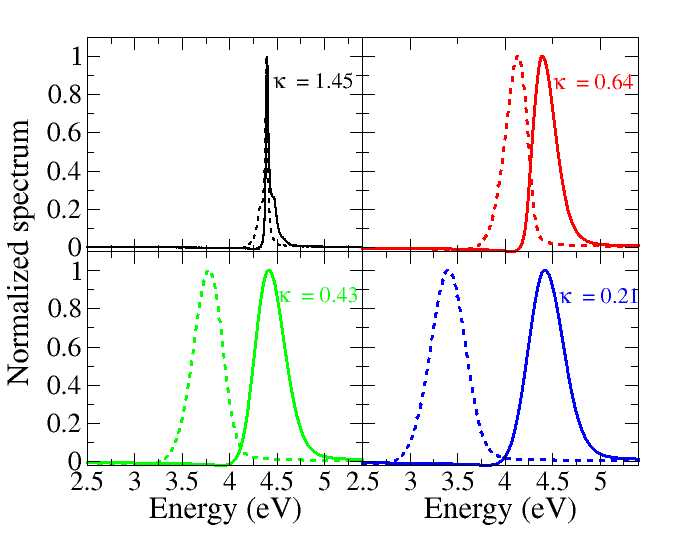}
\caption{Normalized absorption (solid lines) and emission (dashed lines) spectra for the case with a single bright state ($H_{12} = 0$) represented in figure \ref{fig:potentiel}. The panels correspond to the different spectral densities given in figure \ref{fig:density}. $\kappa$ is defined in Eq.(\ref{equ:kappa}). }
\label{fig:spectre}
\end{figure}

(i) The case with a single excited bright state is the basic example for which the behavior is well predicted by the $\kappa$ value. 
When $\kappa >1$, the bath dynamics \textcolor{black}{become} fast and this situation is known as a homogeneous dephasing case leading to Lorentzian profiles for linear absorption or relaxed emission spectra and no Stokes shift. On the contrary, when $\kappa <1$, bath dynamics \textcolor{black}{become} slow, this is the static limit or inhomogeneous case for which the profiles become Gaussian and the maximum Stokes shift is given by $2\lambda $. The four simulations are presented in figure \ref{fig:spectre}. One \textcolor{black}{could observe} the expected behavior passing from no Stokes shift when $\kappa$ = 1.45 to a large shift when $\kappa$ = 0.21 for which $\lambda $ = 0.52 eV. In this example, the relaxation is simply the evolution towards the thermal mixture at the new equilibrium geometry of the excited bright state. This stationary state is obtained in about 150 fs. The ADOs are taken after a propagation of 180 fs.  Generally, the computation of the emission spectrum is more demanding than for the absorption. Convergence of the correlation function requires higher HEOM level (up to level 55 for $\kappa$ = 0.21). Propagation with the TT method has required a small time step of 0.012 fs. The maximum tensor rank remains below 10 with $\epsilon = 10^{-20}$.  \textcolor{black}{The spectra are computed at 298 K. We have verified for the case with $\kappa = 0.21$ that no Matsubara term is necessary.}

\begin{figure}
 \centering
\includegraphics[width =1.05\columnwidth]{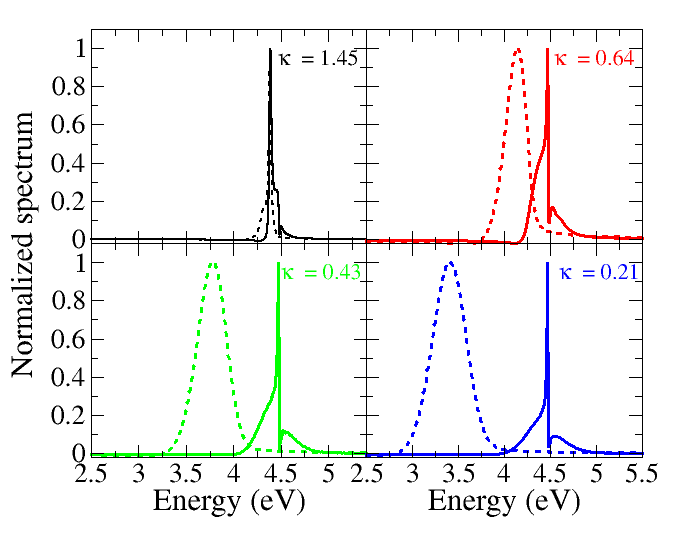}
\caption{Normalized absorption (solid lines) and emission (dashed lines) spectra for the case with a bright state coupled to a dark state (${{H}_{12}}\ne 0$) represented in figure \ref{fig:potentiel}. The panels correspond to the different spectral densities given in figure \ref{fig:density}. $\kappa$ is defined in Eq.(\ref{equ:kappa}). }
\label{fig:spectre2}
\end{figure}

(ii) In the second example presented in figure \ref{fig:spectre2}, dynamics is more complicated since the emission occurs from the eigen vibronic states having a component on the bright state. The propagation duration to reach the asymptotic populations and prepare the ADOs depends on $\lambda$. It is 500 or 600 fs for the different examples. The absorption spectrum is more affected by the nonadiabatic interaction. For the emission spectrum, one recovers the shape corresponding to the relaxed bright state in this example since the second excited state is assumed to be dark. The evolution of the Stokes shift with respect to the $\kappa$ parameter is the same and corresponds to the expected behavior.

These results highlight that tools like HEOM can now predict optical spectra with great precision, given the spectral density. Therefore, attention must now be given to extracting high quality spectral densities to match  experimental results, \textcolor{black}{particulary} in condensed phases where there can be very distinct timescales in the environment, i.e. fast intramolecular motions and slow solvent/lattice/protein reorganisation. The importance of including the latter and a way of obtaining them from first principles was given in Ref.\cite{Dunnett2021bis}.

\textcolor{black}{\subsection {Excitation transfer with discrete bath modes}
In this example, the bath is formed by discrete undamped vibrational modes. We revisit the excitation transfer in a dimer oligothiophene (OT$_4$)-fullerene(C$_{60}$) investigated by I. Burghardt and coworkers \cite{Tamura2011,Burghardt2012,Chenel2014,Burghardt2021}. The excited states may be schematized as in Fig.\ref{fig:potentiel} with $D$ being the excitonic state XT and $B$ becoming the charge transfer (CT) state. We consider the parameters for an inter-fragment distance of 0.35 nm \cite{Chenel2014}.  The energy gap  is 0.07 eV and the electronic coupling is 0.007 eV. The continuous spectral density has been obtained from the difference of the equilibrium position $\Delta q_{eq}$ of the fragment normal modes \cite{Tamura2011}. The reference position of the equilibrated bath is taken at the middle point $\Delta q_{eq}/2$ so that the coupling operator is 
\begin{align}
S=\left( \begin{matrix}
   -0.5 & 0  \\
   0 & 0.5  \\
\end{matrix} \right).
\end{align}
The smooth spectral density \cite{Burghardt2012} is fitted by five two-pole Lorentzians (Eq.(\ref{eq:Johmic})) and displayed in Fig.\ref{fig:dismod}(a). The parameters are given in the supplementary material of Ref.\cite{Chenel2014}. This leads to $K = 10$ artificial decay modes without Matsubara terms at 298 K. We have used the discretization procedure of Ref.\cite{Makri2017} providing unequally spaced modes that correspond to equal fractions of the reorganization
energy. The discrete spectral density at the $N_{disc} = 40$ selected points $k$ ($ \pi c_k^2/(2 \omega_k \Delta(\omega))$ where $\Delta(\omega) $ is the local state density) is shown in Fig.\ref{fig:dismod}(b). Since two baths are involved in the discrete procedure, this corresponds to $K = 80$ modes. Figure \ref{fig:dismod}(c) gives the occupation probability of the XT and CT states when the initial state is XT.  The dashed lines are obtained with the continuous spectral density and the five artificial modes, at level 5 of the hierarchy, with the tolerance parameter $\epsilon = 10^{-10}$ and the maximum rank (rmax) is 20 (see Appendix). This example requires only 3003 matrices and the TT formulation is not more advantageous than the standard one. However, the discrete case may emphasize the utility of the TT propagation when compared with a standard treatment. The corresponding population are drawn in solid lines in Fig.\ref{fig:dismod}(c). We may estimate the total number of elements in the full matrix array. One has $n = 2$, $K = 80$, $L = 5$ leading to ${{N}_{standard}}={{n}^{2}}(K+L)!/\left( K!L! \right) = $ 131,206,068 elements. This is a computationally heavy simulation ($\approx$ 21 Go for the density matrix only) and intractable with our current Fortran implementation. In the TT implementation, one has $n = 2$, $K = 80$, $L = 5$ and $r_{max} = 80$. The number of stored elements might reach a maximum value of ${{N}_{TT}}=({{n}^{2}}+L)*{{r}_{max}}+{{r}_{max}}^{2}*(K-1)*L$ = 2,528,720. The storage is obviously more attractive ($\approx$ 400  Mo). The tolerance is $\epsilon = 10^{-8}$. The norm is well conserved up to about 250 fs. The result is promising but it will be necessary to improve the adaptation of the maximum rank to further extend the performance \cite{Borrelli2021,DunnettChin2021}.}

\begin{figure}
 \centering
\includegraphics[width =1.05\columnwidth]{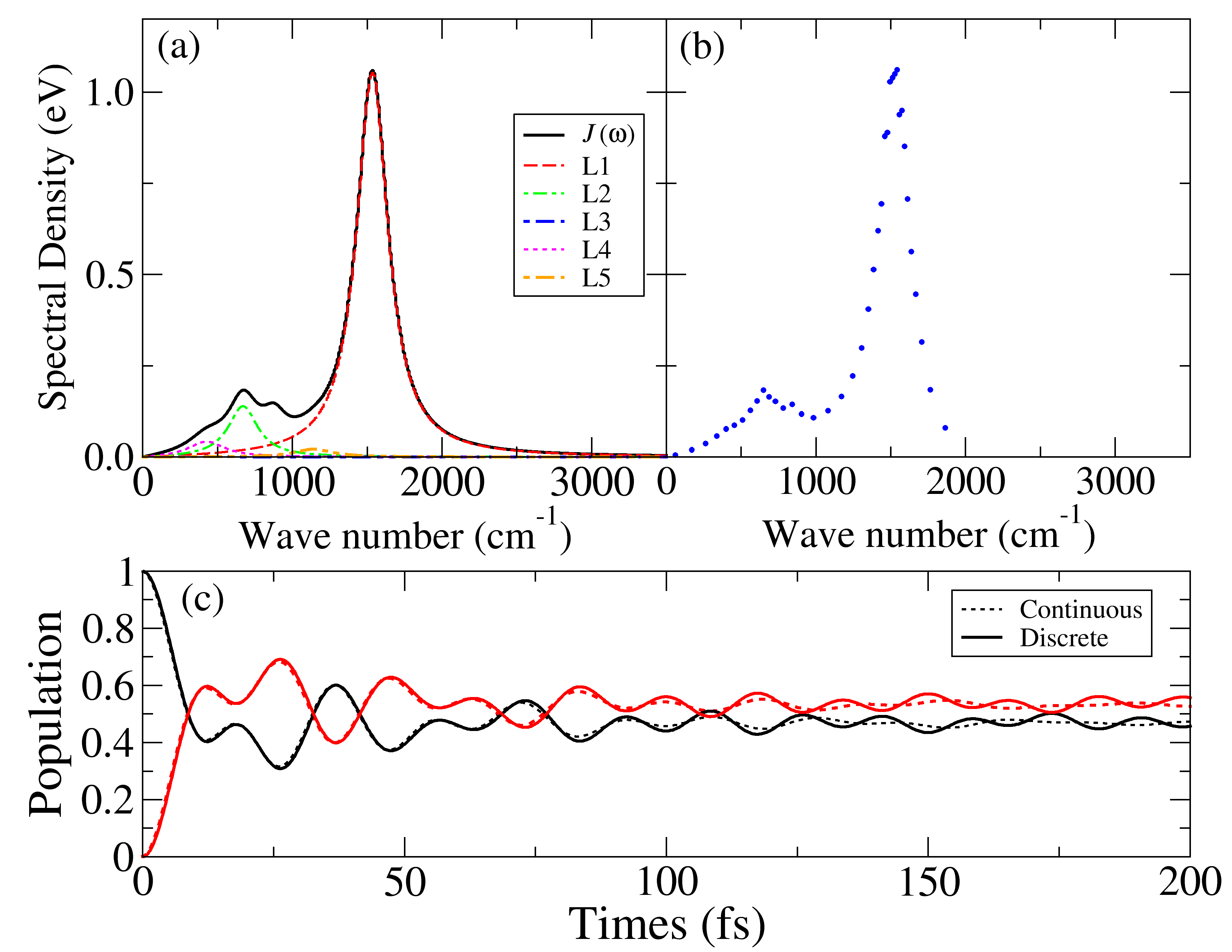}
\caption  {\textcolor{black}{Excitation energy transfer dynamics with a continuous or discretized spectral density by the TT implemetation. (a) : continuous spectral density fitted by five two-pole Lorentzian functions (Eq.(\ref{eq:Johmic})), see Refs.\cite{Burghardt2012,Chenel2014}, (b) : discretization in $N_{disc} = 40$ modes by the Makri procedure (see Ref.\cite{Makri2017}) leading to $K = 2 N_{disc}$ decay modes, (c) : population in the XT (black) and CT (red) states when the initial state is XT. Dashed lines: TT implementation with the continuous spectral density leading to $K = 10$ decay modes without Matsubara terms at 298 K; solid lines : TT implementation with $K = 80$ discrete modes }.}
\label{fig:dismod}
\end{figure}

\section{Conclusion}
 We have given a detailed description of the TT formulation of HEOM and how it is implemented which we hope will be a useful guide to newcomers to the field. This is an exciting development, linking the  widely used HEOM method with rapidly developing advances in tensor networks across physics and theoretical chemistry. \textcolor{black}{One could connect to them the} recent explosion of physical science applications of machine learning, for example: tensor network simulation of multi-environmental open quantum dynamics via machine learning \cite{AspuruGuzik_2017,Zhao2018} and entanglement renormalisation \cite{Schroder2019}. 
 
We have shown \textcolor{black}{three} examples where the TT method allows us to obtain highly accurate results for complex phenomena, such as ‘noisy’ generation of coherences (population to coherence transfer) and the dramatic impact of bath relaxation times on optical spectra. The latter is important, as Stokes shifts are often used as a direct measure of system-bath coupling\textcolor{black}{. This} is clearly only \textcolor{black}{one} part of the story, as the relaxation time is also important\textcolor{black}{. Hence suitable} techniques that can handle multiple timescales (long and short-lived ADOs) are essential. \textcolor{black}{These effects described} above will be even more important in larger multistate systems. \textcolor{black}{The third example simulates an ultrafast excitation transfer by using only undamped decay modes. This example illustrates the efficiency of the TT formulation when compared to the standard method, which would involve a huge number of matrices.} All the applications involve non-Markovian dynamics, either due to long correlation times or high level of hierachy.  

It would be interesting to consider extensions to time-dependent spectroscopies, 2D-spectroscopies \cite{Tanimura_rev_20,Fleming2019,Zhu2011,Tanimura2012} and optical control. Adaptation of the TT formulation for optimal control has already been given in Ref.\cite{Jaouadi2022}. Another important prospect is the consideration of non-harmonic environments, for instance by including some modes in the active system \cite{Chenel2014,MangaudDMP2015,Nazir2016,Gelin2016} and the treatment of fermionic baths \cite{Thoss2021, Lambert2020}.

\section*{\textcolor{black}{Appendix: Numerical implementation of HEOM with ttpy}}
Most of the TT algebra is carried out by the ttpy package developed by Oseledets and coworkers \cite{ttpy}. In this appendix, we show a minimal code to build the system Liouvillian and time-integrate a given system density matrix with initial system-bath factorization in TT format with Python3, Numpy and ttpy packages.

The system Liouvillian is defined as ${\mathcal{L}_{S(ADO)}}=-i(H\otimes {{I}_{n}}-{{I}_{n}}\otimes H)$ where $H$ is the system Hamiltonian, $I_n$ the identity matrix with $n$ the number of system states.   Thus, ${\mathcal{L}_{S(ADO)}}$ is a $n^2 \times n^2$ matrix which can be built from standard numpy functions (\verb!np.eye! returns the identity matrix and \verb!np.kron! the Kronecker product of both matrices) :

\begin{verbatim}
ids = np.eye(n)
Ls = -j * (np.kron(hamil,ids)  \
- np.kron(ids,hamil))
\end{verbatim}
where \verb !ids! is the identity matrix of size $n \times n$ and \verb! j! the imaginary unit.
 To convert this numpy array to a TT format, \verb!tt.matrix! routine performs an approximation of ${\mathcal{L}_{S(ADO)}}$ for a maximal rank (\verb!rmax!) and an accuracy \verb!eps! :
\begin{verbatim}
Ls = tt.matrix(Ls, eps=eps, rmax=rmax)
\end{verbatim}
where \verb !Ls! is the system Liouvillian superoperator.
At this point, we are still only dealing with a representation of an array of $n^2 \times n^2$ dimensions. The HEOM superoperator $\mathcal{L}_{S}$ which spans over the whole Liouville space is defined as $\mathcal{L}_S = \mathcal{L}_{S(ADO)} \otimes \prod_{k''=1}^K I_{n_{HEOM}} $. To avoid memory issues due to the high dimensionality of the array, one must work with the Kronecker products ( \verb!tt.kron!) of ttpy packages instead of the one of numpy (\verb!np.kron!). Indeed, numpy will build the full tensor which can be very large and thus might suffer from the dimensionality curse. 
To carry out this task, a single loop iterates over the number of artificial decay modes $K$ with successive Kronecker products:
\begin{verbatim}
idr = np.eye(nheom)
idr = tt.matrix(idr)
for i in range(K):
    Ls = tt.kron(Ls,idr)
 \end{verbatim}

The total Liouvillian super-operator $\mathcal{L}$ is a sum of several other super-operators, i.e. $\mathcal{L}_{k'}$,  $\mathcal{L}_{k'+}$,  $\mathcal{L}_{k'-}$ (see Eq.(\ref{eq:lioutt})). Addition can be performed directly with the usual algebraic symbol (+) on TT objects. The ttpy library carries out automatically the correct tensor operations.  However, tensor ranks increase at each iteration. Thus, rounding operations are regularly performed to reduce the rank  for a given accuracy and maximum rank with the following command:
\begin{verbatim}
L = L.round(eps,rmax)
\end{verbatim}
where \verb!L!
is the total HEOM Liouvillian super-operator.

The initial vectorized density matrix is defined directly from its cores. The first core is filled with the initial system density matrix. As all auxiliary density matrices are vectors of zeros when assuming system-bath factorization all the other cores are vectors of dimensions $n_{HEOM}$ with only the first index equal to 1.

For a given system density matrix   \verb!rho! 
defined as a numpy array and r the  initial core rank (equal to 1) in this example, we compute \verb!rho! 
the initial density matrix in TT format with the following algorithm:

\begin{verbatim}
rho = np.zeros((n**2), \
      dtype=np.complex128)
rho = np.reshape(rhos,n**2)
cores =[] 
a = np.zeros((1,n**2,r))
a[0,:,0] = rho.copy()
cores.append(a)
vec = np.zeros((nheom), \
      dtype=np.complex128)
vec[0] = 1.
for i in range(K-1):
    a = np.zeros((r,nheom,r))
    a[0,:,0] = vec[:]
    cores.append(a)
a = np.zeros((r,nheom,1))
a[0,:,0] = vec[:]
cores.append(a)
rho = tt.vector().from_list(cores)}
\end{verbatim}

Time-integration is performed with the KSL second-order splitting algorithm \cite{Lubich2014}. For each time step \verb!dt!,
the density matrix in TT format is updated by using the function ksl implemented in tt.ksl routines : 

\begin{verbatim}
rho = tt.ksl.ksl(L,rho,dt)
\end{verbatim}

{By iterating over the desired number of timesteps, we compute the full density matrix at time $t$. In order to extract the system density matrix, projection techniques (by expressing projectors on the full Liouville space in the TT format) or core manipulations can be used. 
}

\bmhead{Acknowledgments}

This work was performed within the French GDR 686 3575 THEMS. We want to dedicate this work to the memory of Christoph Meier and Osman Atabek. Both authors made significant contributions to the research areas described above, such as the Meier-Tannor spectral density and the introduction of the auxiliary matrices. Both have also greatly contributed to many strategies of quantum control in a wide range of processes such as molecular orientation, dynamics of excited states, isomerisation, molecular cooling to cite only few. Dominik Domin is warmly acknowledged for his efficient technical support.

\section*{Declarations}

\begin{itemize}
\item Funding: No funding was received for conducting this study.
\item Conflict of interest/Competing interests: The authors have no relevant financial or non-financial interests to disclose.
\item Ethics approval: Not applicable.
\item Availability of data and materials: This manuscript has no
associated data or the data will not be deposited. [Authors’
comment: The datasets generated during and/or analysed
during the current study are available from the corresponding author on reasonable request.].
\item Authors' contributions: All the authors were involved in the preparation of the
manuscript. All the authors have read and approved the final manuscript.
\end{itemize}

\bibliography{EPJD_2}
\end{document}